\documentclass[useAMS,usenatbib]{mn2e}
\usepackage{pdflscape}
\usepackage{amssymb}
\usepackage{graphicx}
\usepackage{natbib}
\usepackage{lscape}
\usepackage{rotating}
\usepackage{txfonts}
\newcommand{\cpd}{CPD\,$-28^{\circ}$2561}

\newcommand{\ha}{H$\alpha$}
\newcommand{\xmm}{{\sc XMM}\emph{-Newton}}

\newcommand{\loglxlb}{$\log[L_{\rm X}/L_{\rm BOL}]$}
\newcommand{\hei}{\ion{He}{1}}
\newcommand{\heii}{\ion{He}{2}}
\newcommand{\ciii}{\ion{C}{3}}
\newcommand{\niii}{\ion{N}{3}}
\newcommand{\niv}{\ion{N}{4}}
\newcommand{\nv}{\ion{N}{5}}
\newcommand{\civ}{\ion{C}{4}}
\newcommand{\siiv}{\ion{Si}{4}}
\newcommand{\feiv}{\ion{Fe}{4}}

\newcommand{\hst}{{\it HST\/}}
\newcommand\ion[2] {#1\,{\sc{\romannumeral #2}}}

\title[UV and X-ray monitoring of \cpd]{The changing UV and X-ray properties of the Of?p star \cpd \thanks{Based on observations made with \xmm\ (ObsID \# 074018) and and \hst\ (program \# 13629). \xmm\ is an ESA Science Mission with instruments and contributions directly funded by ESA Member states and the USA (NASA). The NASA/ESA Hubble Space Telescope is operated by the Association of Universities for Research in Astronomy, Inc., under NASA contract NAS~5-26555.}}

\author[Naz\'e et al.]{Ya\"el Naz\'e$^1$\thanks{Research Associate FRS-FNRS}\thanks{E-mail: naze@astro.ulg.ac.be}, Jon O. Sundqvist$^{2,3}$, Alex W. Fullerton$^{4}$, Asif ud-Doula$^{5}$, Gregg A. Wade$^{6}$, \and Gregor Rauw$^{1}$, Nolan R. Walborn$^{4}$\\
$^1$ GAPHE, Universit\'e de Li\`ege, Quartier Agora, All\'ee du 6 Ao\^ut 19c, Bat. B5C, B4000-Li\`ege, Belgium\\
$^2$ University of Delaware, Bartol Research Institute, Newark, Delaware 19716, USA\\ 
$^3$ Institut f\"ur Astronomie und Astrophysik der Universit\"at M\"unchen, Scheinerstr. 1, 81679 M\"unchen, Germany\\
$^4$ Space Telescope Science Institute, 3700 San Martin Drive, Baltimore, MD 21218, USA\\
$^5$ Penn State Worthington Scranton, Dunmore, PA 18512, USA\\
$^6$ Department of Physics, Royal Military College of Canada, PO Box 17000, Station Forces, Kingston, ON K7K 4B4, Canada
}
\begin{document}

\date{}
\pagerange{\pageref{firstpage}--\pageref{lastpage}} \pubyear{2015}
\maketitle
\label{firstpage}

\begin{abstract}
The Of?p star \cpd\ was monitored at high energies with \xmm\ and \hst. In X-rays, this magnetic oblique rotator displays bright and hard emission that varies by $\sim$55\% with rotational phase. These changes occur in phase with optical variations, as expected for magnetically confined winds; there are two maxima and two minima in X-rays during the 73d rotational period of \cpd. However, contrary to previously studied cases, no significant hardness variation is detected between minima and maxima, with the exception of the second minimum which is slightly distinct from the first one. In the UV domain, broad-band fluxes remain stable while line profiles display large variations. Stronger absorptions at low velocities are observed when the magnetic equator is seen edge-on, which can be reproduced by a detailed 3D model. However, a difference in absorption at high velocities in the {\ion{C}{4}} and {\ion{N}{5}} lines is also detected for the two phases where the confined wind is seen nearly pole-on. This suggests the presence of strong asymmetries about the magnetic equator, mostly in the free-flowing wind (rather than in the confined dynamical magnetosphere). 
\end{abstract}


\section{Introduction}
The massive O-type star \cpd\ has been known to be peculiar for forty years, as it showed a \heii\,$\lambda$4686 line too strong in comparison to the neighbouring \niii\ emission lines, broad \heii\ lines but sharp \hei\ lines, and strong carbon lines but weak nitrogen lines \citep{wal73,gar77}. However, the causes underlying its strangeness remained hidden until its thorough investigation began, only a few years ago, in the framework of the O and WN survey \citep{bar10,bar14}. At the time, the peculiarities of \cpd\ were found to be similar to those of other members of the class of Of?p stars, despite the absence of strong \ciii\,$\lambda$\,4650 lines (the initial defining characteristic of the Of?p class, see \citealt{wal72,naz08b} for more details). It thus became one of the five known Galactic Of?p stars \citep{wal10}.

Since 2001, it has become clear that the Of?p category gathers massive stars exhibiting several interesting properties. To the presence of peculiar lines or line profiles (e.g. strong \ciii\,$\lambda$\,4650 emission, narrow emissions in Balmer lines, UV line profiles different from those of supergiant Of stars) in their spectra was thus added the presence of periodic variability, notably of their Balmer and \hei\ lines in the optical domain \citep[e.g.][]{naz01,wal04}, but also of their overall brightness \citep[e.g.][]{bar07} and of their (unusually bright) X-ray emission \citep[e.g.][]{naz04,naz07,naz08}. The recurrence timescales range from 7d up to several decades \citep[e.g.][]{wal04,naz06,naz08}. Moreover, and most importantly, strong magnetic fields with global dipolar topologies  \citep[e.g.][]{don06,mar10} were also found to be present in all Galactic Of?p stars. 

These characteristics are understood within the framework of magnetic oblique rotators. For massive stars, a strong magnetic field is able to channel the stellar wind outflow. The effectiveness of the process can be characterized by the wind-confinement parameter \citep{udDoula02}, defined as $\eta_* \equiv B_d^2R_*^2 / ( \dot{M}_{B=0} v_\infty)$, with $R_*$ the stellar radius, $v_\infty$ the wind terminal velocity, $B_d$ the dipole equatorial surface field strength, and $\dot{M}_{B=0}$ the mass-loss rate the star \textit{would have} if it had no magnetic field\footnote{This is essentially the mass feeding rate from the photosphere into the magnetic wind (see e.g. Table \ref{prop}). It is important to realize here that this $\dot{M}_{B=0}$ \textit{is not} the actual mass-loss rate of the star; this actual mass loss is significantly reduced (of the order of $\sim 80$\,\% for $\eta_* \approx 60$, cf. \citealt{udDoula08}) due to the trapped plasma that eventually falls back onto the star.}. The Alfv\'{e}n radius, at which the magnetic and wind energy densities are equal, is $R_{\rm A} \approx \eta_*^{1/4} R_*$ and it is located above the stellar surface if $\eta_* >1$. Below the Alfv\'{e}n radius, the magnetic field is strong enough to channel the radiatively driven wind outflow along closed field lines. The trapped wind plasma channelled along field lines from opposite poles then collides at the magnetic equator. As a result, there is an overdense region around the magnetic equator, which is seen under different angles and/or with different degrees of occultation as the star rotates, explaining the recurrent changes seen throughout the electromagnetic spectrum \citep[for an example of modelling optical variations in this framework, see][]{Sundqvist12}. 

Since massive O stars have high mass-loss rates, the presence of a strong magnetic field has another consequence for those stars: a rapid spindown due to magnetic braking \citep{udd09}, which explains why the Of?p stars rotate slowly. In such a case, rotation is dynamically unimportant and the confined wind plasma falls back to the star by gravity, generating a ``dynamical magnetosphere'' around the star \citep[e.g.][and references therein]{pet13}. 

Indeed, this is the case of \cpd, for which a strong magnetic field was initially detected by \citet{hub11,hub12}. Subsequent spectroscopic and spectropolarimetric monitoring \citep{wad15} further established the star's properties: its (rotation) period is 73.41d, its surface dipolar field $B_d$ amounts to $\sim$2.5\,kG, and its wind magnetic confinement parameter $\eta_*$ is about 100, implying strong confinement. Notably, during one magnetic cycle, the Balmer and \hei\ lines present a double-wave variation, i.e. their emission component is maximum (resp. minimum) twice per period, at $\phi=0.0$ and 0.5 (resp. 0.25 and 0.75). In the framework of magnetic oblique rotators, such an observation implies that both magnetic poles are seen during the stellar rotation, as for the magnetic O9.2\,IV star HD\,57682 \citep{gru12}, implying that the sum of the inclination $i$ and the magnetic obliquity $\beta$ angles is large. Indeed, \citet{wad15} derived $(i,\beta)$ or $(\beta,i)=(90^{\circ},35^{\circ})$. With such a configuration, MHD modelling is able to reproduce the variations of the equivalent width of the \ha\ line \citep{wad15}.

Continuing the work of \citet{wad15}, we present in this paper high-energy monitoring of \cpd, using \xmm\ and \hst\ data. Section 2 presents the observations while the two following sections provide the observational and modelling results in the X-ray and UV domains, respectively. Finally, a summary and conclusions are given in Section 5.

\section{Observations}
\subsection{X-ray data}
\begin{figure}
\includegraphics[width=9cm, bb=15 150 570 700, clip]{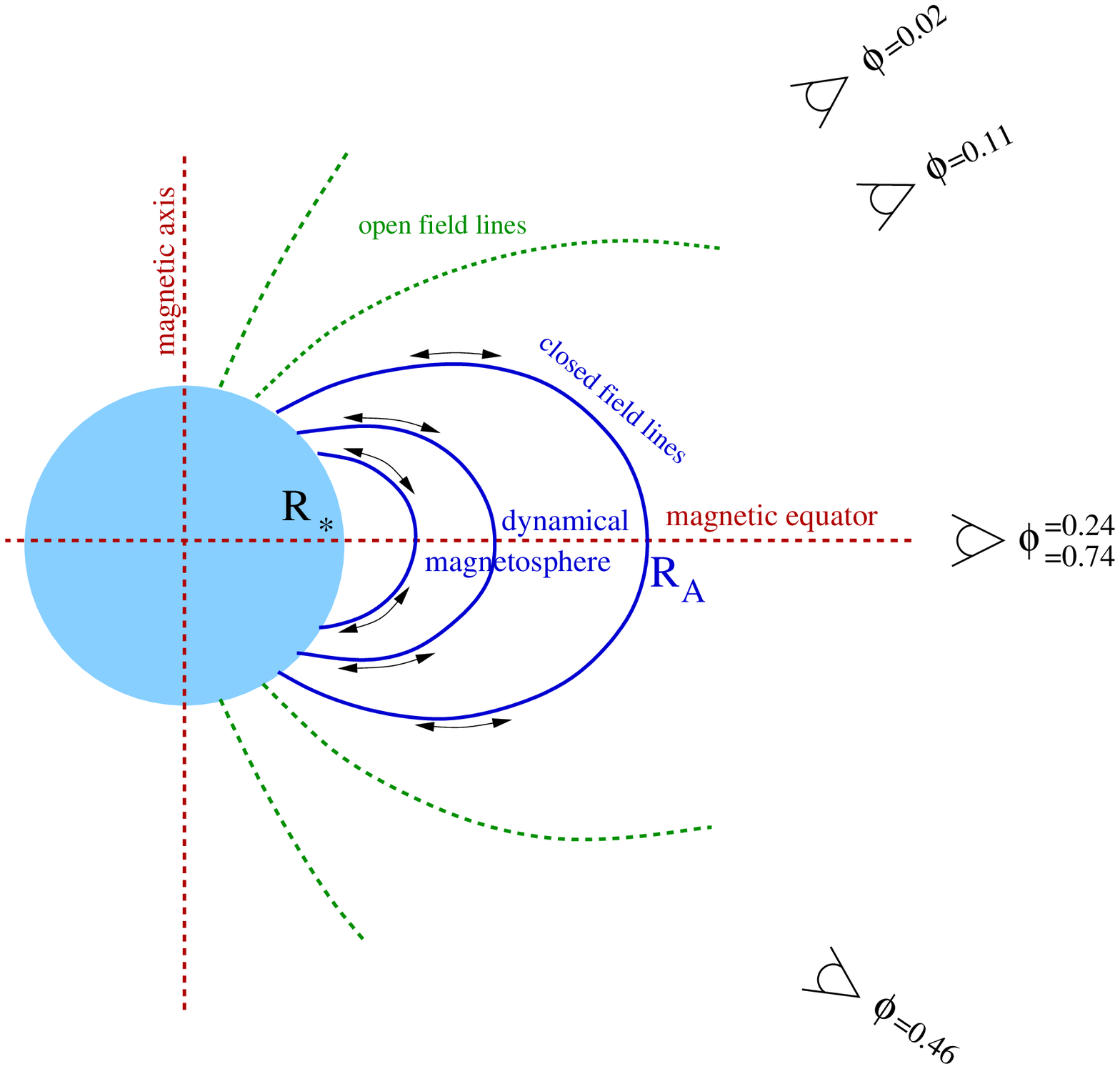}
\caption{Schematic diagram of the system, showing the orientation of the magnetically confined winds of \cpd\ with respect to the observer at the phases of the five \xmm\ observations. Note that the three \hst\ observations were obtained simultaneously with XMM, at $\phi=$ 0.01, 0.25, and 0.49.}
\label{cartoon}
\end{figure}

In 2014, \xmm\ \citep{jan01} observed \cpd\ five times, for 10--20\,ks each time to monitor the changes of the X-ray emission with phase (PI Naz\'e). Table \ref{journal} lists the observation identifiers, the dates, the actual durations (all well below one hundredth in phase), and the phases of the exposures, derived using the ephemeris of \citet[$T_0$=2\,454\,645.49 and $P$=73.41\,d]{wad15}. Note that the first four observations were obtained within the same cycle, but the last one could only be scheduled six months later: this has however not much impact on the phase, considering the small uncertainties on the ephemeris \citep{wad15}. The five exposures sample the two maxima, the two minima, and one intermediate phase in the star rotational cycle (see Fig. \ref{cartoon} for the orientation of the system at these phases). 

\begin{table*}
\centering
  \caption{Journal of observations.}
  \label{journal}
  \begin{tabular}{lcccccc}
  \hline
\# & ObsID & Start dates   & Duration (ks)$^a$ & MJD(mid) & $\phi^b$ & $\alpha^c$ \\
  \hline
\multicolumn{5}{l}{\xmm}\\
1 & 0740180401 (Rev. 2634) & 2014-04-28T07:01:38 & 10.2 & 56775.370 & 29.02 & 55$^{\circ}$ \\
2 & 0740180501 (Rev. 2637) & 2014-05-04T11:58:18 & 4.7  & 56781.631 & 29.11 & 64$^{\circ}$ \\
3 & 0740180601 (Rev. 2642) & 2014-05-14T07:32:40 & 8.9  & 56791.383 & 29.24 & 88$^{\circ}$ \\
4 & 0740180701 (Rev. 2650) & 2014-05-30T11:56:39 & 3.5  & 56807.595 & 29.46 &124$^{\circ}$ \\
5 & 0740180801 (Rev. 2734) & 2014-11-13T18:17:53 & 8.9  & 56974.831 & 31.74 & 92$^{\circ}$ \\
\hline
\multicolumn{5}{l}{\hst}\\
 &   oci7a1010 &  2014-04-27T20:42:35  & 2.110  & 56774.875   & 29.01 & 55$^{\circ}$ \\
 &   oci7a3010 &  2014-05-15T07:58:15  & 2.000  & 56792.344   & 29.25 & 90$^{\circ}$ \\   
 &   oci7a2010 &  2014-06-01T20:11:45  & 1.850  & 56809.852   & 29.49 &125$^{\circ}$ \\
\hline
\end{tabular}
\\
Notes: $^a$ for \xmm, this corresponds to the on-axis actual (i.e., after discarding flares) duration for the pn camera; $^b$ Phases at mid-exposure using the ephemeris of \citet{wad15}; $^c$ Angle between the magnetic axis and the line-of-sight, which can be calculated using $\cos \alpha = \sin \beta \cos (2 \pi \phi) \sin i + \cos \beta \cos i$.
\end{table*}

The \xmm\ data were reduced with the standard Science Analysis System (SAS) software v14.0.0 using calibration files available in December 2014 and following the recommendations of the \xmm\ team\footnote{SAS threads, see \\ http://xmm.esac.esa.int/sas/current/documentation/threads/ }. Only the best-quality EPIC data \citep{str01,tur01} were kept ($PATTERN$ of 0--12 for MOS and 0--4 for pn, note that the source is too faint to have usable RGS data). Large background flares were observed in two exposures (ObsID 0740180501 and 0740180701). The analysis was performed with the times of these flares either retained or discarded. As both analyses yielded similar results, we show below only the results for the cleaned datasets. 

To extract spectra and lightcurves, we must first derive the best source position. To this end, we applied a source detection algorithm on each exposure using the task {\it edetect\_chain} and a likelihood of 10. This was done in the 0.4--10.0\,keV range, considering two energy bands (a soft band corresponding to the 0.4--2.0\,keV energy band and a hard band to the 2.0--10.0\,keV band). This detection run provided the count rate of \cpd\ for each observation (Table \ref{ctrate}) in addition to best-fit positions. To get spectra using the task {\it especget}, source events were then extracted in circular regions centered on these best-fit positions and with a radius of 38\arcsec\ (except for the last exposure, where it was reduced to 33\arcsec\ to avoid CCD gaps). Background events were extracted in a circular region with a radius of 45\arcsec\ (except for the pn data in the last exposure, where it was reduced to 33\arcsec) and centered as close as possible to the target considering crowding and CCD edges. The relative position of background and source was kept the same throughout exposures and cameras (except for the pn data in the last exposure). The EPIC spectra were grouped, using {\it specgroup}, to obtain an oversampling factor of five and to ensure that a minimum signal-to-noise ratio of three (i.e. a minimum of 10 counts) was reached in each spectral bin of the background-corrected spectra. After applying the barycentric correction to event files, EPIC lightcurves were produced for the same regions in the 0.4-2.0\,keV (soft), 2.0-10.0\,keV (hard), and 0.4-10.0\,keV (total) bands and with 1ks time bins. They were then corrected using {\it epiclccorr} to get equivalent on-axis, full PSF count rates corrected for (known) photon losses. In addition, as previous experience shows, very large errors and wrong estimates of the count rates are avoided by discarding bins displaying effective exposure time $<$50\% of the time bin length. We also checked that the raw source+background lightcurves and the background-corrected lightcurves of the source yield similar results. Finally, it may be noted that the source is not bright enough to present pile-up. 

\subsection{UV data}
\subsubsection{Photometry}
In parallel to its X-ray telescopes, \xmm\ possesses a small optical/UV telescope called the Optical Monitor (OM, \citealt{mas01}) which aims at observing faint sources. With the OM, we obtained UV photometry of \cpd\ in both the UVM2 filter (centered on 2310\AA\ with a width of 480\AA) and the UVW2 filter (centered on 2120\AA\ with a width of 500\AA). For each filter, five short subexposures of 1--2ks were taken, except for the third \xmm\ observation where only one long exposure of 5ks was obtained per filter. The data were acquired in image mode for all exposures except the last one (which used only one filter, UVM2, and the image-fast mode). No photometry in other filters could be obtained as \cpd\ is too bright in the other bands. In fact, the target is so bright (about 400 counts per seconds in the UVM2 filter and 200 cts\,s$^{-1}$ in the UVW2 filter) that coincidence loss correction may introduce errors in the results (particularly in fast mode) - see section 4.1 for details. As recommended by the SAS team, we reduced these data using the task {\it omichain} and {\it omfchain} (the latter for the fast mode only). Note that \cpd\ has no close neighbour nor any close straylight feature which could contaminate its photometry.

\subsubsection{Spectroscopy}
In addition to the \xmm\ data, we also obtained three ultraviolet spectra of {\cpd} with the Space Telescope Imaging Spectrograph (STIS) on board the {\it Hubble Space Telescope} ({\hst}) under the auspices of the Joint {\hst}/XMM-Newton Observing Proposals program. Table~{\ref{journal}} provides a journal of these observations, which constitute {\hst} GO Program~13629 (PI: Naz\'e). They were obtained nearly simultaneously with the first, third, and fourth \xmm\ observations, i.e. they sample the two maxima and one minima according to the ephemeris of \citet{wad15}.  

\begin{figure*}
\includegraphics[width=8.5cm]{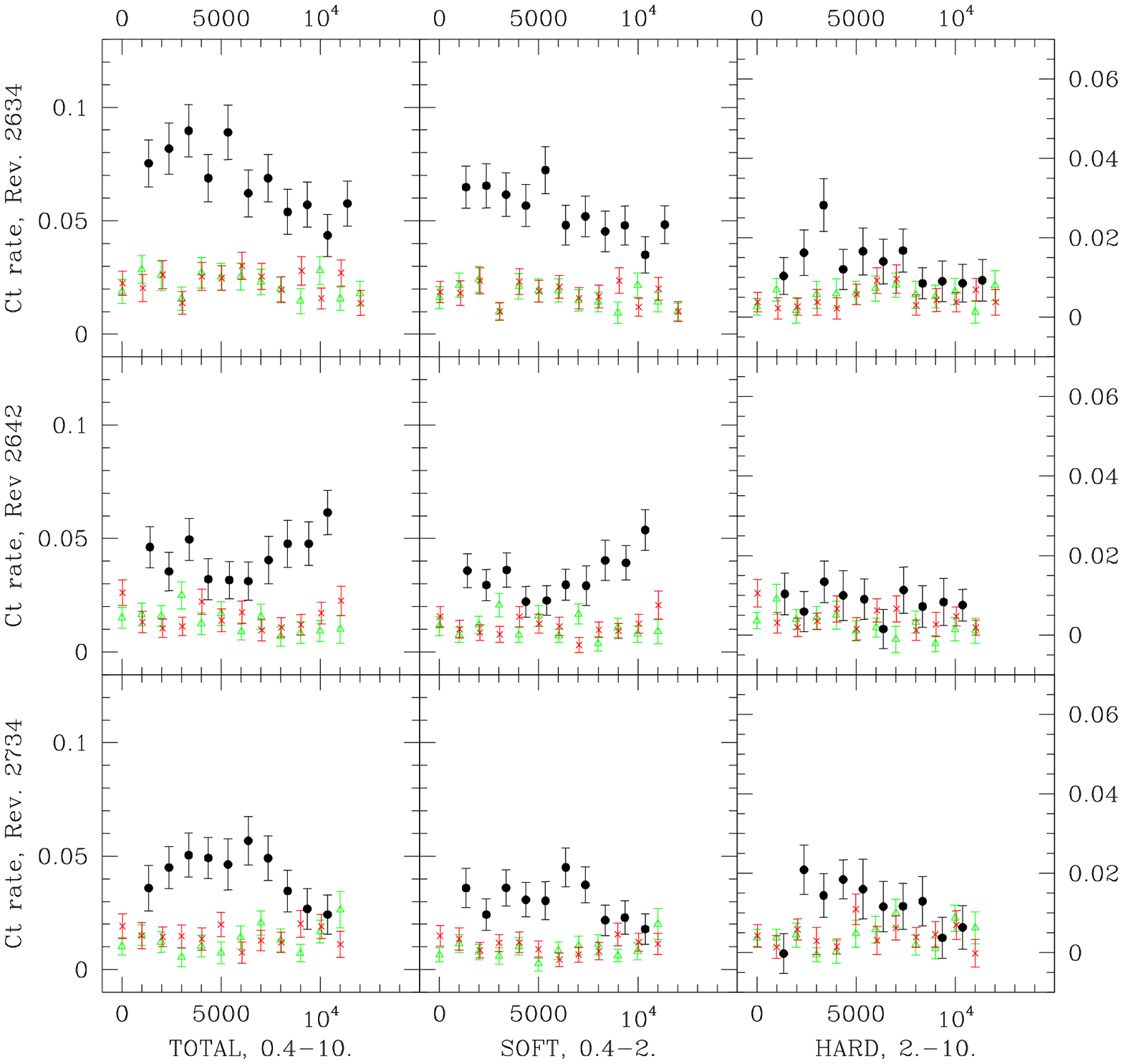}
\includegraphics[width=8.5cm]{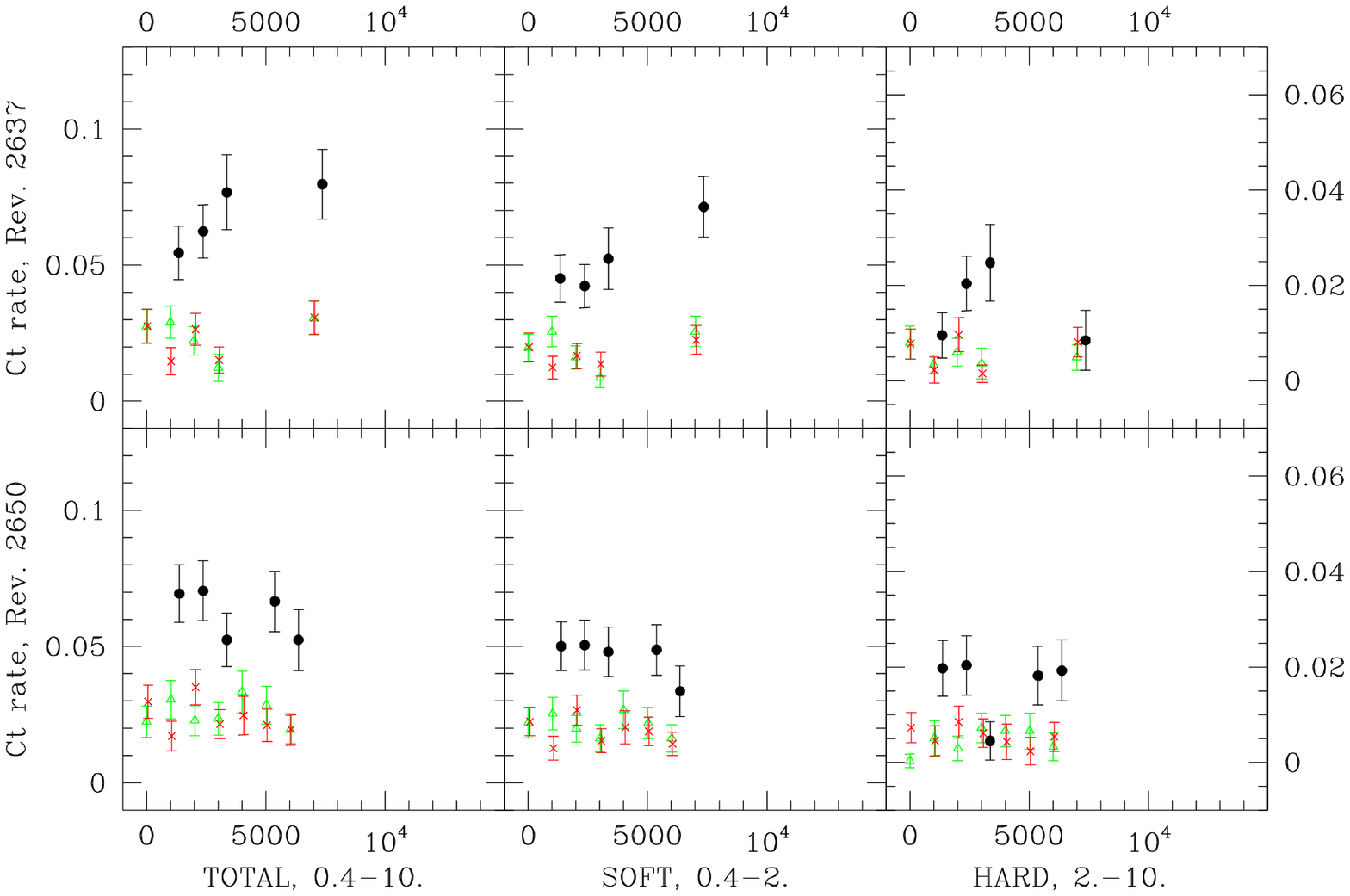}
\caption{Intra-pointing lightcurves derived from the \xmm\ observations. The x-axis gives the time, in seconds, elapsed since the beginning of the observation. Black dots correspond to pn data, green triangles to MOS1 data, and red crosses to MOS2 data. Each row corresponds to one observation and displays the lightcurves in three different bands (from left to right: total, soft, hard).}
\label{lcx}
\end{figure*}

All the STIS spectra were obtained in the same manner.  The observing sequence consisted of a standard spectroscopic acquisition in the F28X50OII aperture, an ``auto-wavecal"  exposure of the internal Pt-Cr/Ne wavelength calibration lamp, and a single exposure of {\cpd} through the $0.2\arcsec \times 0.2\arcsec $ aperture with the standard ACCUM mode of the far-ultraviolet MAMA photon-counting detector. Spectra were produced by the E140M grating, which provided wavelength coverage from {1144~\AA} (order 129) to {1729~\AA} (order 86) with a resolving power of $\sim$46\,000. Since successive visits had a decreasing visibility window, the exposure time varied from 2.11 to 1.85 ks, but in all cases produced a signal-to-noise ratio of about 30 per pixel near the blaze maximum of the most sensitive orders.

The spectra were uniformly processed with version 3.3 (2013 October 03) of the CALSTIS pipeline, which included correction for detector nonlinearity, dark current subtraction, flat field correction, determination and application of corrections to the wavelength zero point, 1-D spectral extraction of each order, application of dispersion solutions, photometric calibration, and scattered-light corrections. Additional effort was devoted to extracting order 86, which contains the {\ion{N}{4}\,$\lambda$\,1718} resonance line.  Although this order falls completely on the MAMA detector and is well separated from adjacent orders, it has not been routinely extracted by CALSTIS for the past few observing cycles.  This exclusion is evidently due to small shifts in the location of the order, which push the regions used to model the background and scattered light off the detector and preclude the standard correction for these effects.  Since these corrections are less significant for orders that are well-separated, we developed a simplified method of tracing the order, estimating the local background, and extracting the net count rates that uses modified versions of the current reference files.  

As a final step, the extracted, calibrated orders for each observation were merged into a single spectrum.

\section{The X-ray emission of \cpd}
In strongly magnetic massive stars, the collision between the wind flows from both hemispheres produces a bright and hard X-ray emission \citep[and references therein]{bab97,udd14}. Depending on the geometry of the magnetosphere, the X-ray emitting regions in such a magnetic oblique rotator may be regularly occulted by the star as the system rotates, which yields periodic X-ray variations locked in phase with other periodically variable quantities, such as the recorded magnetic field and the intensities of optical lines. In the next subsections, we examine the properties of the X-ray emission of \cpd, to determine its intensity, variability, and relationship to other observational quantities.

\subsection{Lightcurves}

\begin{figure}
\includegraphics[width=9cm]{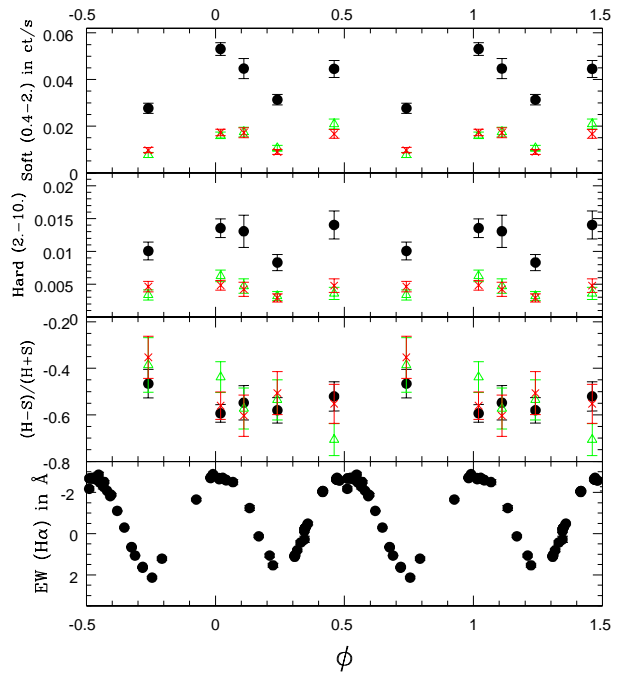}
\caption{Evolution with phase of the X-ray count rates in the different bands and of the X-ray hardness ratio of \cpd, compared to the \ha\ variations recorded in the optical domain \citep{wad15}. Black dots correspond to pn data, green triangles to MOS1 data, and red crosses to MOS2 data.}
\label{ctratefig}
\end{figure}

We searched for X-ray variability in the light curves of \cpd\ that occurs on short (intra-pointing) and long (inter-pointig) time scales. Considering first the intra-pointing variability, we applied a series of $\chi^2$ tests (with three different null hypotheses: constancy, linear variation, quadratic variation, as for $\zeta$\,Pup, \citealt{naz13}) to the derived lightcurves (Fig. \ref{lcx}). We further compared the improvement of the $\chi^2$ when increasing the number of parameters in the model (e.g. linear trend vs constancy) by means of F-tests. Adopting a significance level (SL) of 1\%, we found that the lightcurves are never significantly variable, but also that the pn lightcurves in the total and soft bands of the first, third, and last exposures are significantly better fit by linear or parabolic trends than by a constant. Indeed, these lightcurves (left panels of Figure \ref{lcx}) show obvious maxima with flux dropping on each side or minima with flux increasing on each side. The larger noise in MOS data (due to redirection of half of their flux into the RGS) prevents us from confirming these results. While more precise X-ray data are needed for a more complete investigation, the detected variations can already be easily understood qualitatively. In fact, the short-term variability only reflects the phase-locked variations of the star with the rotational period. As we will see below, X-ray and optical emissions vary simultaneously: as the phases of these observations correspond to extrema in the optical spectrum and X-ray emission, this explains the shape of the observed lightcurves (which are simply close-up on maxima or minima). 

The similarity in behaviour between X-ray and optical domains is evident from the analysis of the inter-pointing variability. When phased with the ephemeris of \citet[][see Table \ref{ctrate} and Fig. \ref{ctratefig}]{wad15}, the count rates derived from the source detection algorithm show maxima at $\phi=0.0$ and 0.5, and minima at $\phi=0.25$ and 0.75, with the intermediate phase ($\phi=0.11$) yielding intermediate results. It may be noted that the total and soft count rates of the second maximum ($\phi\sim0.5$) are slightly smaller than those of the first one ($\phi\sim0$), but this difference is small (less than 3$\sigma$, see Table \ref{ctrate}). As for $\theta^1$\,Ori\,C \citep{gag05,ste05} and HD\,191612 \citep{naz07,naz10}, the X-ray flux variations are thus clearly phased with the stellar rotation period, as demonstrated by the simultaneous minima in X-ray and optical emissions (Fig. \ref{ctratefig}). In addition, the amplitude of the X-ray changes is also similar: the maximal count rates of \cpd\ are $\sim$60\% larger than the minimal ones, comparable to what is observed for $\theta^1$\,Ori\,C, HD\,191612, and Tr16-22 \citep{naz14}. 

In addition to brightness changes, varying spectral shapes are observed for the X-ray emission of $\theta^1$\,Ori\,C, HD\,191612, and Tr16-22 \citep{naz14}. The presence of such X-ray hardness changes is probably related to a temperature stratification in the magnetospheric plasma, with the warm and hot regions being occulted by different amounts. The situation appears different for \cpd, however, as no significant variations of hardness are present over the first four exposures and only a slightly harder emission is found in the last observation. The stratification in the magnetosphere of \cpd\ thus appears less pronounced than in other magnetic O-stars.

The variations in X-ray brightness are generally considered to be  mostly due to occultation of the X-ray emitting regions by the stellar body as the star rotates. To test to what extent such an occultation can account for X-ray variation in \cpd, we employed a simple toy model wherein the X-ray emission region is an optically-thin ring-like region in the magnetic equatorial plane with negligible thickness but varying widths. One can then derive the expected degree of occultation as a function of geometry of the region, phase, inclination of the star ($i$), and magnetic obliquity ($\beta$). We tested different geometries: a ``disk'' from $R_*$ to $R_A$ and a ring with a radius restricted to $R=(R_A-R_*/2)\pm R_* /2$, which better mimics the X-ray emitting regions seen in MHD simulations \citep[e.g.][]{udDoula08}. We also tested several possibilities for the brightness distribution within the X-ray emitting region: uniform brightness, brightness varying as $r^{-2}$, and brightness varying as $r^{-4}$. For the parameters ($i$, $\beta$, $R_A$) of the three magnetic stars with known geometry and X-ray monitoring ($\theta^1$\,Ori\,C, HD\,191612, and \cpd), the ratio between the maximum and minimum fluxes predicted by this simple model amounts to 110\%--130\%, and it does not change when considering thicker ring-like regions. This predicted ratio is smaller than the observed ratios ($\sim$140--160\%), suggesting that the simple occultation of optically-thin regions near $R_A$ is not the dominant process at the origin of the X-ray variations. However, if such emitting regions are located closer to the photosphere, the agreement is better. But it is not obvious how X-rays could be produced near the surface as this process requires high wind shock velocities which can only occur further out. Clearly, a more sophisticated modelling is required, and will be investigated in the future using fully self-consistent 3D MHD models. 

\begin{table*}
\centering
  \caption{X-ray photometry of \cpd. The hardness ratio is defined as $HR=(H-S)/(H+S)$, with $S$ and $H$ the count rates in the soft (0.4--2.0\,keV) and hard (2.0--10.0\,keV) bands, respectively.}
  \label{ctrate}
  \begin{tabular}{lc|cccc|cc|cc}
  \hline
\# & $\phi$ & \multicolumn{4}{|c|}{pn (cts\,s$^{-1}$)} & \multicolumn{2}{|c|}{MOS1 (cts\,s$^{-1}$)} & \multicolumn{2}{|c}{MOS2 (cts\,s$^{-1}$)}\\
   &        & Total & Soft & Hard & $HR$ & Total & $HR$ & Total & $HR$\\
\hline
1 & 0.02 & 0.067$\pm$0.003 & 0.053$\pm$0.003 & 0.0135$\pm$0.0014 & $-$0.59$\pm$0.04 & 0.0223$\pm$0.0016 & $-$0.44$\pm$0.07 & 0.0221$\pm$0.0016 & $-$0.56$\pm$0.06 \\
2 & 0.11 & 0.058$\pm$0.005 & 0.045$\pm$0.004 & 0.0131$\pm$0.0024 & $-$0.55$\pm$0.07 & 0.0222$\pm$0.0023 & $-$0.57$\pm$0.09 & 0.0213$\pm$0.0024 & $-$0.60$\pm$0.09 \\  
3 & 0.24 & 0.040$\pm$0.003 & 0.031$\pm$0.002 & 0.0083$\pm$0.0012 & $-$0.58$\pm$0.05 & 0.0137$\pm$0.0014 & $-$0.54$\pm$0.09 & 0.0118$\pm$0.0012 & $-$0.51$\pm$0.09 \\
4 & 0.46 & 0.059$\pm$0.004 & 0.045$\pm$0.004 & 0.0140$\pm$0.0021 & $-$0.52$\pm$0.06 & 0.0245$\pm$0.0023 & $-$0.71$\pm$0.07 & 0.0214$\pm$0.0021 & $-$0.55$\pm$0.08 \\
5 & 0.74 & 0.038$\pm$0.003 & 0.028$\pm$0.002 & 0.0101$\pm$0.0014 & $-$0.47$\pm$0.06 & 0.0110$\pm$0.0014 & $-$0.39$\pm$0.12 & 0.0143$\pm$0.0014 & $-$0.35$\pm$0.09 \\
\hline
\end{tabular}
\end{table*}

\begin{figure*}
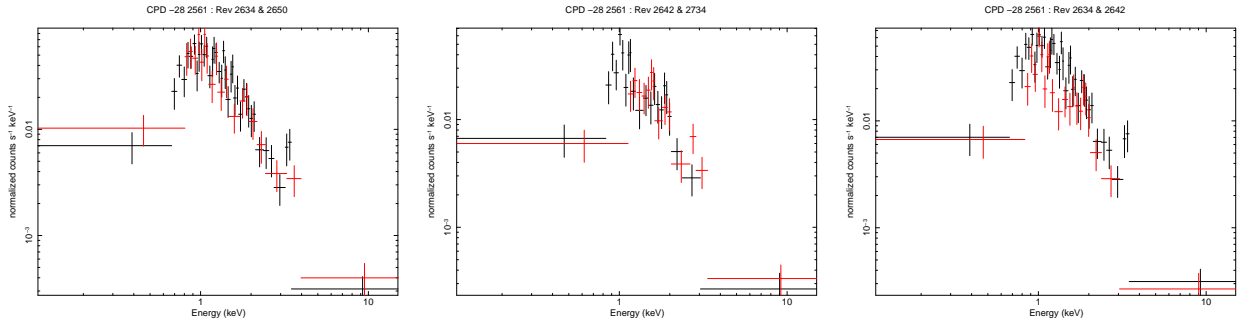

\includegraphics[height=5.5cm,angle=270]{spec4.ps}
\includegraphics[height=5.5cm,angle=270]{spec5.ps}
\includegraphics[height=5.5cm,angle=270]{spec2.ps}
\caption{X-ray spectra of \cpd\ in the \xmm\ observations. Only pn data are shown for clarity. {\it Left:} Comparison between the pn data taken at the two maxima ($\phi=0.02$ in black, $\phi=0.46$ in red). {\it Middle:} Comparison between the pn data taken at the two minima ($\phi=0.24$ in black, $\phi=0.74$ in red). {\it Right:} Comparison between the pn data taken at the first maximum ($\phi=0.02$ in black) and the first minimum ($\phi=0.24$ in red).}
\label{xspec}
\end{figure*}

\subsection{Spectra}
\begin{figure}
\includegraphics[width=9cm, bb=25 180 585 440, clip]{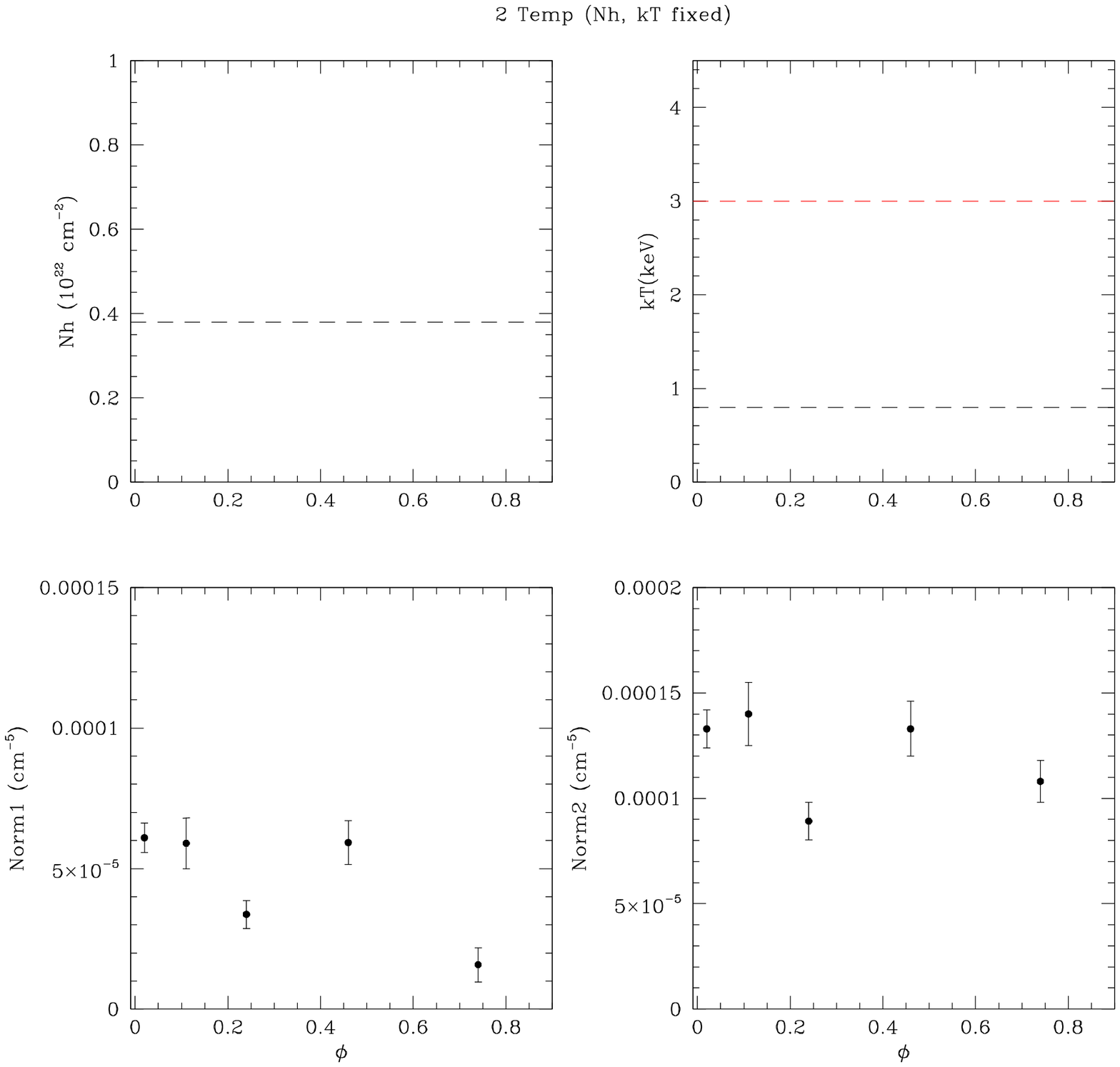}
\includegraphics[width=9cm]{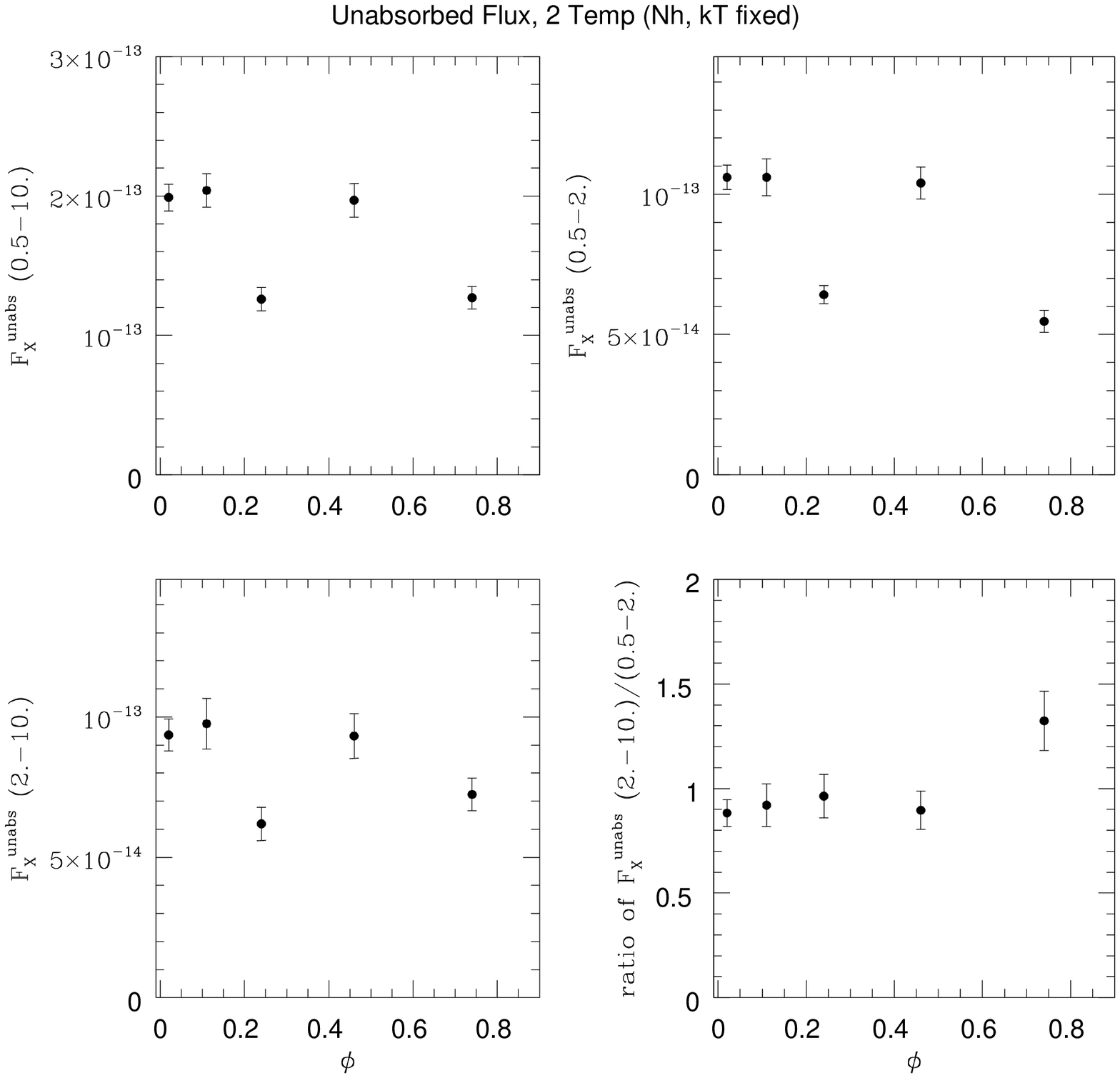}
\caption{Evolution with phase of the best-fit normalization factors and ISM absorption corrected fluxes (for the model with fixed absorption and two fixed temperatures, see Table \ref{fit}). }
\label{fitfig}
\end{figure}

To get more detailed information, we then turned to the X-ray spectra (Fig. \ref{xspec}). The five observations were considered separately, as variations exist (see previous section), but the fitting procedure was kept the same. For each observation, all EPIC data (MOS1, MOS2, and pn) were fitted simultaneously within Xspec v12.8.2 using absorbed optically-thin thermal plasma models, i.e. $wabs \times phabs \times \sum apec$, with solar abundances \citep{aspl}. The first absorption component is the interstellar column, fixed to $1.8\times10^{21}$\,cm$^{-2}$ (a value calculated using the conversion formula $5.8\times10^{21}\times E(B-V)$\,cm$^{-2}$ from \citealt{boh78} and the color excess of the star $E(B-V)$=0.31, estimated from $B-V$=0.04), while the second absorption represents additional (local) absorption. For the emission component, we proceeded in several steps. First, we considered two thermal components (one thermal component was not enough to achieve a good fitting). In the resulting fits, the additional absorption and temperatures did not significantly differ amongst datasets. We then fixed them to $3.8\times10^{21}$\,cm$^{-2}$, 0.8\,keV and 3.0\,keV, respectively, for a last set of fits. Second, to reproduce (to first order) a multi-temperature plasma, we fitted a series of four emission components with temperatures fixed to 0.2, 0.6, 1.0, and 4.0\,keV as used in the global X-ray analysis of magnetic hot stars by \citet{naz14}. This allows us to directly compare with the global survey results. In these fits, the additional absorption was not observed to significantly vary, hence we decided to fix it at the average value of $6.6\times10^{21}$\,cm$^{-2}$ for another set of fits. Spectral parameters derived by these different fitting procedures are provided in Table \ref{fit} (see also Fig. \ref{fitfig}). It should be noted that the different types of models yield similar $\chi^2$ and similar results, within the errors. 

\begin{table*}
\centering
  \caption{Best-fit parameters for the fit of X-ray (EPIC) spectra. }
  \label{fit}
  \begin{tabular}{lccccccccccc}
  \hline
\multicolumn{10}{l}{\it model with 2 thermal components}\\
\# & $\phi$ & $N_{\rm H}$ & $kT_1$ & $norm_1$ & $kT_2$ & $norm_2$ & $\chi^2$ (dof) & $F_{\rm X}^{obs}$ (tot.) &  (soft) & (hard) & $F_{\rm X}^{unabs}$ \\ 
& & (10$^{22}$\,cm$^{-2}$) & (keV) & (10$^{-5}$\,cm$^{-5}$) & (keV) & (10$^{-4}$\,cm$^{-5}$) &  & \multicolumn{4}{c}{(10$^{-14}$\,erg\,cm$^{-2}$\,s$^{-1}$)} \\ 
\hline
1 & 0.02 & 0.34$\pm$0.29 & 0.76$\pm$0.12 & 5.16$\pm$6.76 & 2.71$\pm$0.39 & 1.40$\pm$0.14 & 0.96(77) & 16.2$\pm$1.4 & 7.45$\pm$0.60 & 8.70$\pm$0.91 & 19.6\\
2 & 0.11 & 0.35$\pm$0.25 & 0.86$\pm$0.12 & 5.08$\pm$5.52 & 2.71$\pm$0.71 & 1.42$\pm$0.23 & 0.93(33) & 16.4$\pm$2.6 & 7.46$\pm$1.00 & 8.92$\pm$1.90 & 19.7\\
3 & 0.24 & 0.26$\pm$0.26 & 0.96$\pm$0.18 & 2.55$\pm$2.54 & 3.08$\pm$0.98 & 0.84$\pm$0.13 & 1.05(40) & 10.5$\pm$1.0 & 4.50$\pm$0.41 & 5.97$\pm$1.30 & 12.5\\
4 & 0.46 & 0.66$\pm$0.34 & 0.58$\pm$0.35 & 17.3$\pm$194. & 3.85$\pm$4.21 & 1.20$\pm$0.62 & 1.17(42) & 17.4$\pm$3.3 & 7.50$\pm$0.90 & 9.94$\pm$2.64 & 20.8\\
5 & 0.74 & 0.31$\pm$0.35 & 1.03$\pm$0.40 & 1.86$\pm$3.27 & 3.94$\pm$2.47 & 0.96$\pm$0.23 & 1.29(32) & 12.1$\pm$2.2 & 3.95$\pm$0.75 & 8.17$\pm$1.80 & 13.7\\
\hline
1 & 0.02 & 0.38$^f$ & 0.8$^f$ & 6.10$\pm$0.52 & 3.$^f$ & 1.33$\pm$0.09 & 0.95(80) & 16.6$\pm$0.8 & 7.39$\pm$0.30 & 9.18$\pm$0.56 & 19.9\\
2 & 0.11 & 0.38$^f$ & 0.8$^f$ & 5.90$\pm$0.90 & 3.$^f$ & 1.40$\pm$0.15 & 0.88(36) & 17.0$\pm$1.0 & 7.45$\pm$0.46 & 9.56$\pm$0.88 & 20.4\\
3 & 0.24 & 0.38$^f$ & 0.8$^f$ & 3.37$\pm$0.50 & 3.$^f$ & 0.89$\pm$0.09 & 1.02(43) & 10.6$\pm$0.7 & 4.52$\pm$0.23 & 6.07$\pm$0.58 & 12.6\\
4 & 0.46 & 0.38$^f$ & 0.8$^f$ & 5.93$\pm$0.78 & 3.$^f$ & 1.33$\pm$0.13 & 1.16(45) & 16.4$\pm$1.0 & 7.28$\pm$0.40 & 9.15$\pm$0.78 & 19.7\\
5 & 0.74 & 0.38$^f$ & 0.8$^f$ & 1.58$\pm$0.61 & 3.$^f$ & 1.08$\pm$0.10 & 1.22(35) & 11.0$\pm$0.7 & 3.98$\pm$0.28 & 7.11$\pm$0.57 & 12.7\\
\hline
\multicolumn{10}{l}{\it model with 4 thermal components of temperatures 0.2, 0.6, 1.0, and 4.0\,keV}\\
\# & $\phi$ & $N_{\rm H}$ & $norm_1$ & $norm_2$ & $norm_3$ & $norm_4$ & $\chi^2$ (dof) & $F_{\rm X}^{obs}$ (tot.) &  (soft) & (hard) & $F_{\rm X}^{unabs}$ \\ 
& & (10$^{22}$\,cm$^{-2}$) & (10$^{-4}$\,cm$^{-5}$) & (10$^{-5}$\,cm$^{-5}$) & (10$^{-5}$\,cm$^{-5}$) & (10$^{-4}$\,cm$^{-5}$) &  & \multicolumn{4}{c}{(10$^{-14}$\,erg\,cm$^{-2}$\,s$^{-1}$)} \\ 
\hline
1 & 0.02 & 0.86$\pm$0.17 & 9.88$\pm$9.70   & 8.16$\pm$6.36 & 7.48$\pm$3.61 & 9.96$\pm$1.30 & 0.94(77) & 16.9$\pm$1.0 & 7.62$\pm$0.50 & 9.27$\pm$0.85 & 20.5 \\
2 & 0.11 & 0.76$\pm$0.22 & 7.25$\pm$10.4   &   0.$\pm$5.05 & 11.6$\pm$5.60 & 10.1$\pm$2.10 & 0.96(33) & 17.2$\pm$5.5 & 7.51$\pm$4.70 & 9.74$\pm$1.90 & 20.8 \\
3 & 0.24 & 0.50$\pm$0.20 & 0.98$\pm$17.2   &   0.$\pm$2.40 & 4.85$\pm$1.72 & 7.20$\pm$1.42 & 1.07(40) & 11.1$\pm$2.1 & 4.51$\pm$1.48 & 6.59$\pm$0.90 & 13.2 \\
4 & 0.46 & 0.83$\pm$0.36 & 7.56$\pm$22.5   & 16.0$\pm$10.3 & 1.00$\pm$5.39 & 11.7$\pm$1.80 & 1.15(42) & 17.6$\pm$3.0 & 7.54$\pm$2.80 & 10.1$\pm$1.50 & 21.1 \\
5 & 0.74 & 0.34$\pm$0.43 &0.0(unconstrained)&   0.$\pm$6.30 & 2.01$\pm$2.87 & 9.63$\pm$1.23 & 1.26(33) & 12.2$\pm$1.3 & 3.95$\pm$0.90 & 8.25$\pm$0.75 & 13.8 \\
\hline
1 & 0.02 & 0.66$^f$ & 3.71$\pm$1.96 & 6.24$\pm$3.63 & 6.14$\pm$1.99 & 10.6$\pm$1.10 & 0.95(78) & 17.2$\pm$1.0 & 7.54$\pm$0.28 & 9.69$\pm$0.79 & 20.9 \\
2 & 0.11 & 0.66$^f$ & 4.44$\pm$2.53 &   0.$\pm$4.77 & 10.5$\pm$2.80 & 10.4$\pm$1.80 & 0.94(34) & 17.4$\pm$3.2 & 7.48$\pm$1.43 & 9.88$\pm$1.26 & 20.9 \\
3 & 0.24 & 0.66$^f$ & 2.32$\pm$1.69 & 0.03$\pm$2.96 & 6.29$\pm$1.58 & 6.77$\pm$1.16 & 1.05(41) & 10.9$\pm$2.1 & 4.53$\pm$1.44 & 6.38$\pm$0.85 & 13.0 \\
4 & 0.46 & 0.66$^f$ & 2.08$\pm$3.44 & 12.6$\pm$6.90 & 1.64$\pm$3.61 & 11.7$\pm$1.60 & 1.13(43) & 17.6$\pm$2.0 & 7.48$\pm$1.28 & 10.2$\pm$1.20 & 21.2 \\
5 & 0.74 & 0.66$^f$ & 0.00$\pm$1.95 & 3.24$\pm$2.01 & 2.78$\pm$2.41 & 9.32$\pm$1.30 & 1.27(33) & 12.0$\pm$2.9 & 4.01$\pm$1.80 & 8.03$\pm$1.05 & 13.6 \\
\hline
\end{tabular}
\\
Notes: $^f$ means fixed parameter ; for similarity with other papers, energy bands are here defined as 0.5-2.0\,keV (soft), 2.0-10.0\,keV (hard), and 0.5-10.0\,keV (total) ; $F_{\rm X}^{unabs}$ is the flux after correction for ISM-absorption in the total band. The relative error on the latter quantity is assumed to be similar to the relative error of the observed flux, though this does not take into account the error coming from the choice of model.\\
\end{table*}

The fitting results agree with what was derived from the count rates: the ISM-absorption corrected fluxes appear larger by $\sim$55\% at maxima ($\phi=0.02$, 0.11, and 0.46) compared to minima ($\phi=0.24$ and 0.74), while the hardness of the X-ray emission (traced by the flux ratio and the relative importance of normalization factors, or average temperature\footnote{Note however that the average temperature is not always a consistent estimate of plasma hardness \citep{naz14}.}) increases only in the last observation ($\phi=0.74$). While the two maxima are quite similar, it clearly appears that the two minima (at phases 0.24 and 0.74) are different with respect to spectral shape (see also Fig. \ref{xspec}). The second minimum corresponds to the last observation, taken 6 months (i.e. about 2 cycles) after the first four exposures. A sudden change in magnetospheric structure would be extremely unlikely, as the optical spectra present an excellent periodicity: no large change in the confined wind behaviour was detected over the $>$25 cycles covered by these data \citep{wad15}. The change observed in X-rays may thus rather be related to the asymmetrical structure of the magnetosphere. Indeed, the \ha\ emission line displays an obvious radial velocity shift and a profile skew change between the two maxima (phases 0.0 and 0.5), which might plausibly be linked to, e.g., an off-centered dipole. This asymmetry between the two poles will have an effect on the channelling, hence on the X-ray production. No model of that asymmetry exists, however, to compare with the data.

Finally, \cpd\ has an average \loglxlb $\sim-5.8\pm0.1$ and a flux hardness ratio (=$F_{\rm X}^{unabs}$(hard)/$F_{\rm X}^{unabs}$(soft)) close to one. Compared to other magnetic O-stars \citep{naz14}, \cpd\ thus appears as slightly brighter (by 0.4\,dex with respect to average) and harder (in the survey, most magnetic O-stars have a flux hardness ratio of about 0.3, with only Plaskett's star and $\theta^1$\,Ori\,C rivalling \cpd). This is not due to the value of the local absorption, which is quite typical of magnetic O-stars. Considering the stellar properties of  \cpd\ (Table \ref{prop}), the semi-analytic X-ray modelling by \citet{udd14} yields an X-ray luminosity of $\log(L_{\rm X})\sim 33.2$, or a \loglxlb $\sim-5.7$, for \cpd, considering a 10\% efficiency as found adequate in both MHD models and observational surveys \citep{naz14}. The small difference with the observed value may result from the imperfect knowledge of the wind velocity or initial ($B=0$) mass-loss rate. The theoretical value is also close to that derived for HD\,191612 \citep{naz14} which is expected as this star presents stellar properties similar to those of \cpd. It may be noted, however, that these objects differ slightly on the observational side as HD\,191612 has \loglxlb $\sim-6.05$ and a flux hardness ratio of $\sim$0.3.

\subsection{Comparison with other X-ray observatories}
Few other X-ray data of \cpd\ exist. The star is reported as 1RXS\,J075552.8$-$283741 in the {\it ROSAT} faint source catalog, with a count rate of 0.027$\pm$0.013\,cts\,s$^{-1}$. Folding our best-fit models through the {\it ROSAT} response matrices results in an expected count rate of 0.004--0.006\,cts\,s$^{-1}$, compatible with the reported value at $<2\sigma$, considering the (large!) errors. A single {\it Suzaku} observation was serendipitously taken on the exact same date as the first \xmm\ exposure. Its analysis was presented by \citet{hub14}. While there is a broad similarity of results with our first \xmm\ observation (no short-term variability, presence of a hard component, large \loglxlb), it must be noted that the \xmm\ data are of much higher quality: for example, the lightcurves of the background in the total band appears at count rates ten times lower than the source lightcurves for \xmm, whereas that factor is only about two for {\it Suzaku} (see Fig. 10 of \citealt{hub14}); many spectral bins in {\it Suzaku} actually are lower limits, especially at lower energies (their Fig. 11). This explains why the spectral model of \citet{hub14} has a reduced $\chi^2$ of 5.2 when used on \xmm\ data. Compared to \citet{hub14}, we may also further note that we use the more recent $apec$ model for the thermal X-ray emission modelling as it uses the latest atomic/ionic properties (the older $mekal$ resulting in different spectral parameters), and that we considered absorption in addition to the interstellar one, as is known to be necessary for O-stars: this explains the modelling difference between that paper and this work. 

\begin{table}
\centering
\caption{Summary of the properties of \cpd. The first ten lines are copied from \citet{wad15}. The mass-loss rate $\dot{M}_{\rm B=0}$ given in this table was calculated using the formula of \citet{Vink00}: this value (slightly different from the value reported in \citealt{wad15}) was used for the calculation of $\eta_*$ and $R_{\rm A}$ reported in the last two lines. Note that the bolometric luminosity places the star at $\sim$8\,kpc, that $i$ and $\beta$ are interchangeable, and that the terminal velocity, assumed by \citet{wad15}, appears compatible with UV observations (see next section). }
\label{prop}
\begin{tabular}{l|ll}
\hline
Parameter & Value \\
\hline
$T_{\rm eff}$ (K) & 35 000 $\pm$ 2000  \\
$\log g$ (cgs) & 4.0 $\pm$ 0.1     \\
R$_{*}$ (R$_\odot$) & 12.9 $\pm 3.0$  \\
$\log (L_*/L_\odot)$ & 5.35$\pm 0.15$ \\
$v\sin i$ (km\,s$^{-1}$) & $\lesssim 80$   \\
$P_{\rm rot}$ (d) & $73.41\pm 0.05$ \\
$v_{\infty}$ (km\,s$^{-1}$) & 2400  \\
$B_{\rm d}$ (G) & $2600\pm 900$ \\
$i$ ($\degr$) & $35\pm 3$ \\
$\beta$ ($\degr$) & $90\pm 4$ \\
$\log \dot{M}_{\rm B=0}$ (M$_{\odot}$\,yr$^{-1}$) & $-6.4$  \\
$\eta_*$ &  50-2300 (230)  \\
$R_{\rm A}$ ($R_*$) & 2.8-6.8 (4.2) \\
\hline
\end{tabular}
\end{table}

\section{The UV emission of \cpd}
\subsection{OM Photometry}

\begin{figure*}
\includegraphics[width=8.5cm]{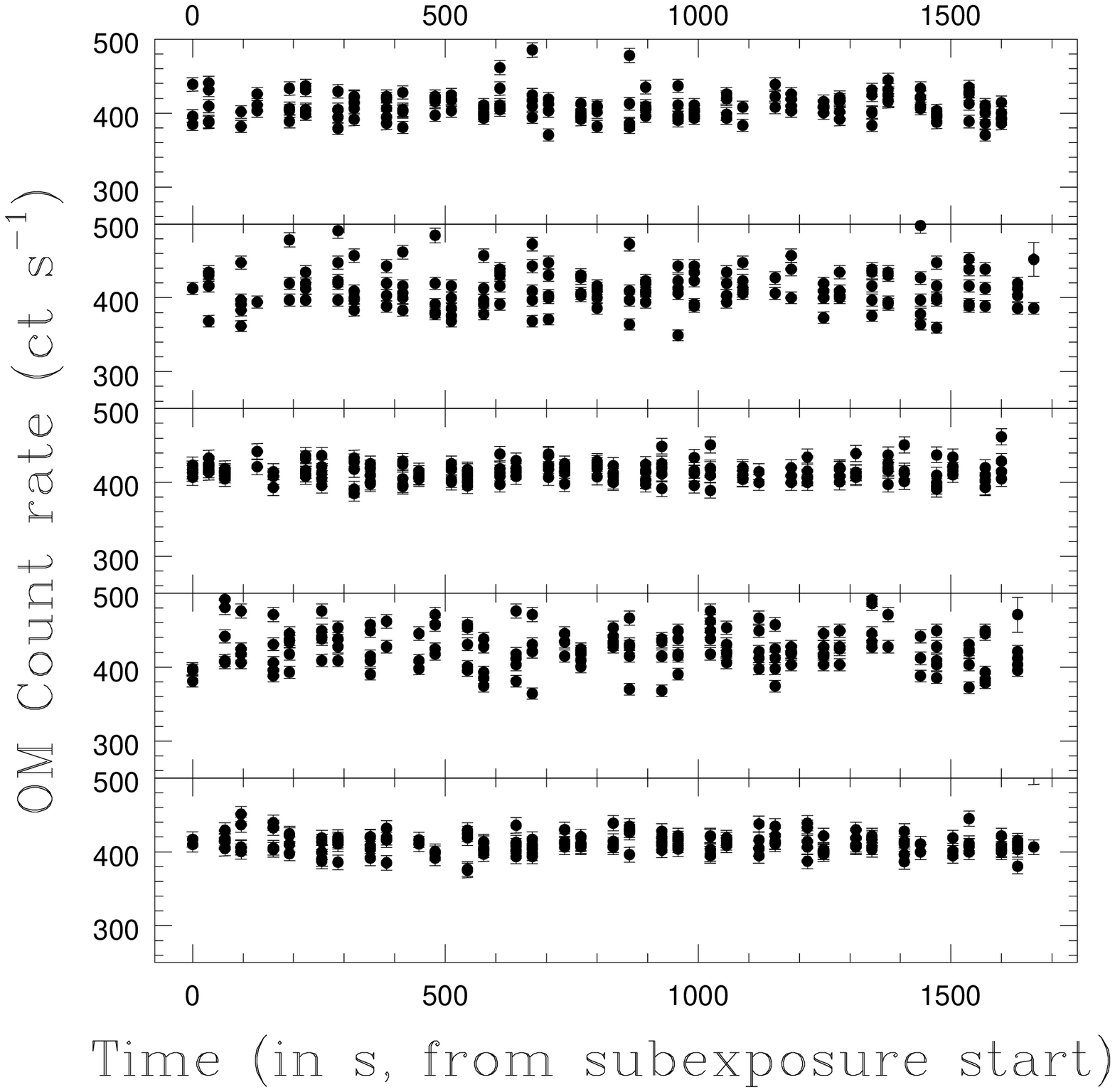}
\includegraphics[width=8.5cm]{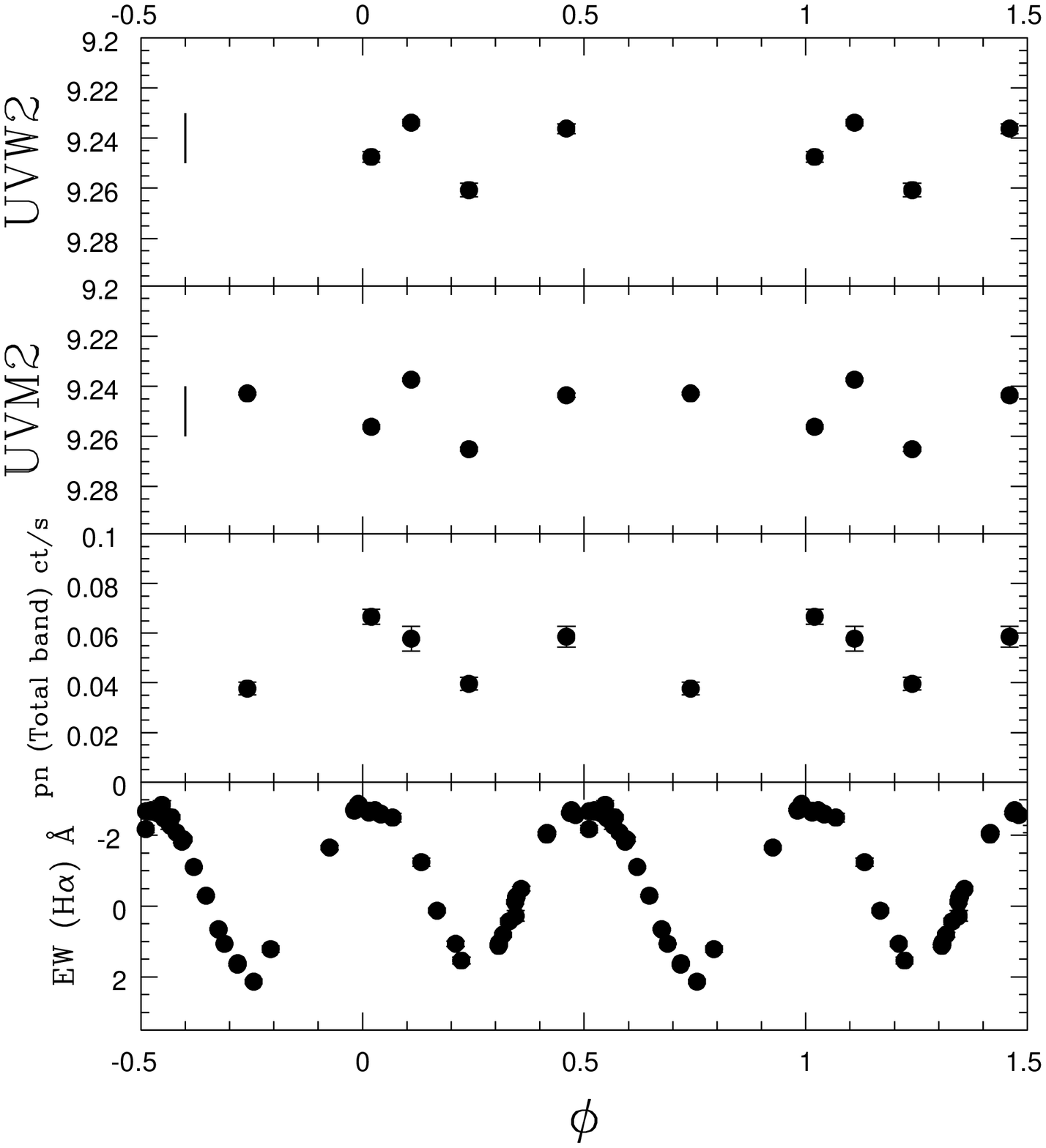}
\caption{{\it Left:} UV lightcurves in UVM2 filter for the five subexposures of the last \xmm\ observation ($\phi=0.74$, taken in fast mode). {\it Right:} Evolution with phase of the UV magnitudes of \cpd, compared to the X-ray count rate (pn, total band) and \ha\ variations. The vertical bars at the left of the top two panels indicate the typical calibration error, to be added to the source specific errors (see text).}
\label{lcuv}
\end{figure*}

We first examine UV photometry, taken with the OM telescope onboard \xmm. For all observations except the third one, there are five OM subexposures per filter (see e.g. left panel of Fig. \ref{lcuv}), so that the intra-pointing variability can be investigated. In addition, the last observation has OM data taken in fast mode, ensuring one measure every 0.5s within each subexposure to study any rapid changes in the UV emission of the target. However, since \cpd\ is bright, the use of bins with 50--100s duration is recommended for fast mode data to diminish the effect of coincidence losses (see Section 2.2.1) so we could not study variability on shorter timescales than these bin durations. 

A $\chi^2$ test on the source count rates was performed: it did not reveal any significant variability when comparing the stellar brightness measured in the five subexposures of a single observation. The same was true when we examined the individual (fast-mode) lightcurves of the last observation, except for one of them - but fast mode is the most affected by coincidence losses so that this single potential variability detection needs confirmation. This implies that changes in the UV emission rarely occur on short timescales (from tens of seconds to a few ks). 

Turning to even longer timescales by examining inter-pointing variability, Table \ref{om} provides the average UV magnitudes derived for each of the OM observations after combining all subexposures (see also Fig. \ref{lcuv}). At first sight, one might conclude that significant variability is present. If real, the UV changes would apparently lack any phase coherence with respect to the optical and X-ray variations (Fig. \ref{lcuv}). However, it must be kept in mind that the errors in Table \ref{om} are underestimated as they do not take into account the systematic errors. This is particularly true for bright sources such as \cpd, since the former errors are very small in that case so that the latter errors largely dominate. Fortunately, stability monitoring of the OM calibration and data reduction system was made and it found measurements of standard (bright) stars to be stable by 2\% or 0.02\,mag (I. de La Calle, private communication). The observed ``variations'' of \cpd\ are of similar amplitude (see vertical lines in Fig. \ref{lcuv}), hence casting doubt on their actual presence. We may thus conclude that the emission of \cpd\ over broad UV bands remains stable. This is confirmed by the perfect overlap of the fluxed UV spectra (see Fig. \ref{specuv}) but it does not prevent line profile changes in that domain to be present, however, as we will see in the next section.

\begin{table}
\centering
  \caption{OM photometry of \cpd.}
  \label{om}
  \begin{tabular}{lcccc}
  \hline
\# & $\phi$ & ObsID & UVW2 & UVM2\\
\hline
1 & 0.02 & 0740180401 & 9.2475$\pm$0.0022 & 9.2562$\pm$0.0009 \\
2 & 0.11 & 0740180501 & 9.2338$\pm$0.0012 & 9.2374$\pm$0.0006 \\
3 & 0.24 & 0740180601 & 9.2607$\pm$0.0027 & 9.2651$\pm$0.0009 \\
4 & 0.46 & 0740180701 & 9.2362$\pm$0.0019 & 9.2436$\pm$0.0009 \\
5 & 0.74 & 0740180801 &                   & 9.2429$\pm$0.0006 \\
\hline
\end{tabular}
\end{table}

\subsection{STIS Spectroscopy}

\begin{figure*}
\includegraphics[width=18cm]{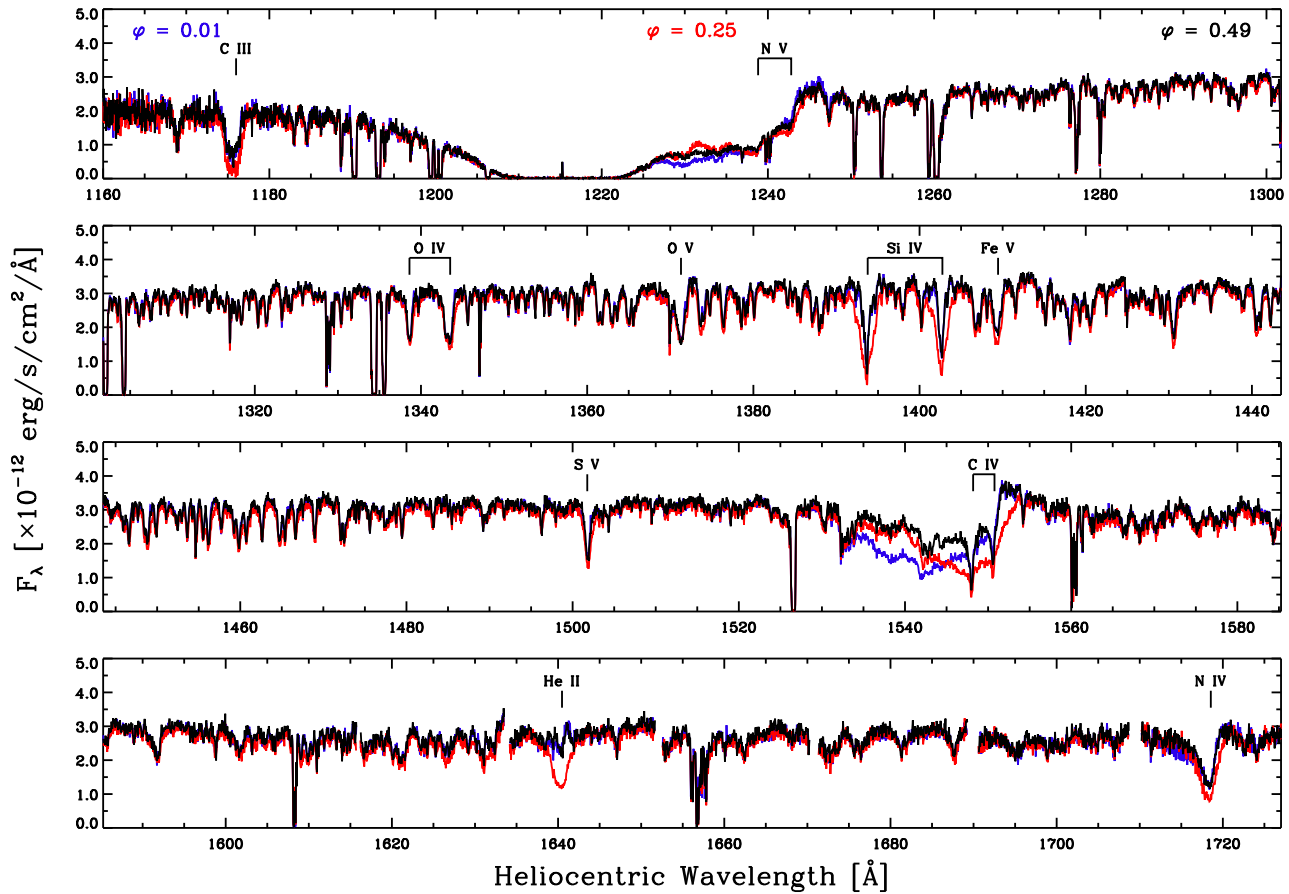}
\caption{STIS spectra of {\cpd} at three phases (0.01 in blue, 0.25 in red, and 0.49 in black), with identification of the key transitions.}
\label{specuv}
\end{figure*}

Figure~\ref{specuv} illustrates the three STIS spectra of {\cpd} that were obtained at quarter-cycle intervals of the magnetic variation; see Fig. \ref{cartoon} for the orientation of the observer with respect to the magnetospheric structure inferred by \citet{wad15}. 

As discussed by \citet{wad15}, the spectral type of \cpd\ varies between $\sim$O6.5  at its ``high state'' (which corresponds to $\phi=0.0$ and 0.5, when \ha\ exhibits emission maxima) and O8 at its ``low state'' (i.e. $\phi=0.25$ and 0.75). These classifications are based on the changes in the ratio of {\ion{He}{1}\,$\lambda$4471} to {\ion{He}{2}\,$\lambda$4542}, with the understanding that the low state probably gives a better indication of the general properties of the stellar photosphere because the {\ion{He}{1}} lines are partially filled with emission during the high state. The line strengths in the UV spectra are broadly consistent with a ``late O'' classification, but are similarly characterized by significant abnormalities. In particular, the presence of {\ion{O}{5}\,$\lambda$\,1371 is atypical for the temperature classes indicated by the optical spectra, since it usually appears at similar strengths only at the earliest types (i.e., O2--O5). Its occurrence, together with the excessive strength of the {\ion{N}{4}\,$\lambda$\,1718} P~Cygni profile indicates the presence of additional highly ionized gas.  Another abnormality concerns the {\ion{Si}{4}} doublet. As usual in spectra of mid-to-late O dwarfs, its components at phases 0.01 and 0.49 are absorption features with a strong interstellar contribution. However, the doublet becomes dramatically broader at $\phi = 0.25$ (\ha\ minimum), which is anomalous when compared to the morphological trends exhibited by non-magnetic O-type stars. 

Finally, the most striking peculiarities occur in the P~Cygni profiles of the {\ion{N}{5}} and {\ion{C}{4}} doublets, which are significantly different at all three phases.  These profiles are highlighted in Fig. \ref{uv_spectra2}, together with the {\ion{Si}{4}} doublet and {\ion{He}{2}\,$\lambda$\,1640} as well as spectra of {\ion{He}{2}\,$\lambda$4686} and {\ha} obtained at similar phases by \citet{wad15}. At all phases, the morphology of the P~Cygni wind profiles is quite different from that expected for a spherically symmetric wind, primarily because of the weakness of the emission lobe.

The biggest changes in the UV spectra are seen at $\phi=0.25$, corresponding to \ha\ minimum - a time at which the line of sight to a distant observer passes through the material that collects near the magnetic equator. Enhanced absorption is visible in many lines at this phase, notably the {\ion{Si}{4}} doublet as already mentioned but also ``photospheric'' features like {\ion{Fe}{5}\,$\lambda$\,1409} as well as diagnostics of circumstellar material like the {\ion{C}{3}}\,$\lambda$\,1176 multiplet and the {\ion{N}{4}\,$\lambda$\,1718} line. The {\ion{He}{2}\,$\lambda$\,1640} line exhibits enormously enhanced absorption at this phase, which is all the more noticeable because the line is either completely filled in by emission or displays a weak P~Cygni profile at both \ha\ maxima. The increased density along the line-of-sight at $\phi=0.25$ is evidently accompanied by changes in the velocity structure that reduce the amount of forward-scattered emission along the line of sight: this explains the enhanced absorption at low velocities in the P~Cygni absorption trough of the {\ion{N}{5}} and {\ion{C}{4}} doublets, which weakens the emission lobe and shifts the observed emission peak to longer wavelengths. The simultaneous enhancement of lines that arise from many different species and stages of ionization confirms that increased density is the primary cause of the increased optical depths, rather than changes in the ionization fractions.

Although the P~Cygni profiles of the {\ion{N}{5}} and {\ion{C}{4}} doublets show the presence of high-velocity material at all phases, the absorption at high velocities is greatest near $\phi$=0.01 when the fast polar wind is exposed (see Fig.~\ref{cartoon} and Fig.~\ref{uv_spectra2}).  However, when the wind emerging from the opposite hemisphere is visible at phases near $\phi$=0.49, the absorption trough is substantially reduced at all velocities even though the emission lobe is essentially identical in appearance. These differences, together with the similar strength but reversal in skew exhibited by  {\ha} and {\ion{He}{2}\,$\lambda$4686} at the two maxima (see Fig.~\ref{uv_spectra2} and \citealt{wad15}), indicate a significant asymmetry in the magnetosphere when viewed from these different perspectives. 

\begin{figure}
       \resizebox{\hsize}{!}  {\includegraphics[]{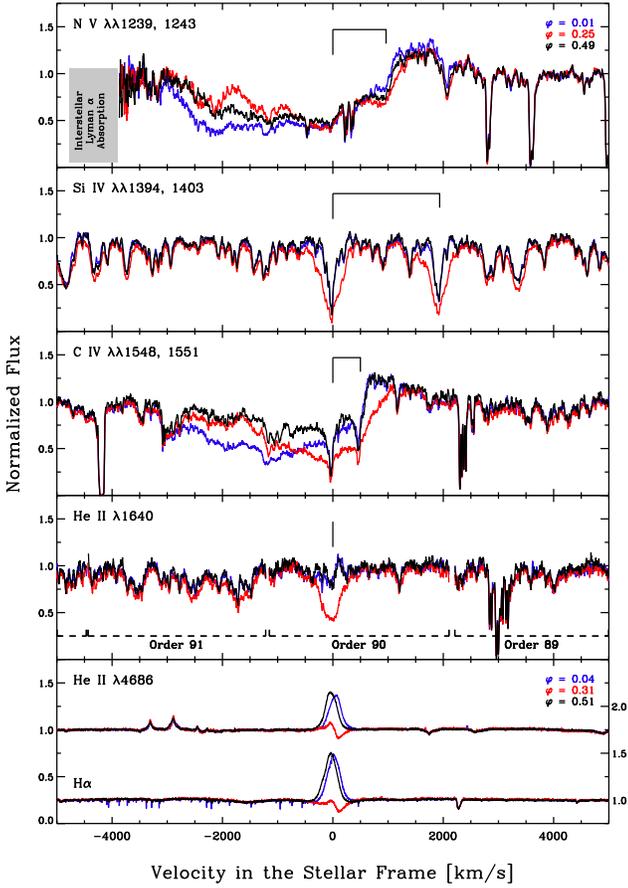}}
   \caption{Normalized UV lines serving as diagnostics of the wind and magnetosphere for three phases (0.01 in blue, 0.25 in red, and 0.49 in black). For comparison, the variations of {\ion{He}{2}}\,$\lambda$4686 and H$\alpha$ at similar phases are shown at the same scale.  The optical data were obtained with CFHT/ESPaDOnS \citep{wad15}.}
   \label{uv_spectra2}
\end{figure}

\subsubsection{Comparison with other magnetic massive stars}
Although the oblique magnetic rotator model provides a useful conceptual basis for understanding the variability of stars like {\cpd}, it is rich in parameters that determine how different diagnostics appear and behave. For example, the detailed behavior of a resonance line in a magnetic O-type star depends on the mass flux from the surface of the underlying star, the stellar rotation rate, the strength and geometry of the magnetic field, the competition between radiative, magnetic, and centripetal forces on the ionization balance in both the magnetosphere and the free-streaming wind, and the observer viewing aspect $\alpha$ (see last column of Table \ref{journal}). Despite the diversity of this parameter space and the current difficulty to get UV spectra, some patterns of behaviour for ultraviolet diagnostics are starting to emerge as more phase-resolved spectroscopy becomes available. 

UV spectroscopy is currently available for three Of?p stars \citep[HD\,108, HD\,191612, and \cpd,][and this work]{mar12,mar13}, HD\,57682 \citep{gru09}, and the young magnetic O-star $\theta^1$\,Ori\,C \citep{sta96,smi05}. 

The Of?p stars HD\,191612 and HD\,108 display weaker P~Cygni profiles (i.e. less blueshifted absorption and more redshifted absorption) of strong UV lines when the \ha\ emission is minimum (hence equatorial regions are seen edge-on) while weaker UV lines, having no blueshifted absorption, show only stronger low-velocity absorptions at this phase \citep{mar12,mar13}. For HD\,108, the strong lines were \nv\,$\lambda$\,1240, \siiv\,$\lambda$\,1400, \civ\,$\lambda$\,1550, and \niv\,$\lambda$\,1718, while the \feiv\ forest fell in the weak line category. For HD\,191612 as for \cpd, \siiv\,$\lambda$\,1400 changed category, hence changed behaviour. This dichotomy could be qualitatively reproduced by MHD simulations \citep[see Fig. 6 of][]{mar13}. Strong lines, such as those of \civ, are sensitive even to the high-speed wind flowing out of the poles, despite its lower density compared to the confined plasma of the equatorial regions. Therefore, such lines display absorption over a large range of velocities. They can be broadly described as P~Cygni profiles, which become stronger when the system is seen pole-on (and the free-flowing polar wind comes into view). When the equatorial region is seen edge-on, the line-of-sight absorbing column mostly traces the dynamical magnetosphere region characterized by low velocities, so that the P~Cygni profiles appear globally weaker. In contrast, intrinsically weak lines are not sensitive to the outflowing wind, so they exhibit less or even negligible absorption at high velocities and display a simple absorption profile located at low velocities. This absorption logically increases when the dense and slowly-moving confined winds of the equatorial regions enter the line-of-sight. In this framework, \cpd\ appears as both similar and different. Indeed, purely photospheric lines are constant while low-velocity absorptions are indeed stronger at $\phi$=0.25, when the dense equatorial regions enter the line-of-sight. However, the changes in \civ\ (in particular its different profiles at the two maxima) and \heii\ are novel.

Outside the Of?p category, the reported behaviours are quite varied. The magnetic O-type star HD\,57682 seems perfectly in line with Of?p stars, as its \siiv\ doublet displays stronger and broader absorptions at \ha\ minimum \citep[see Fig. 4 of ][ considering ephemeris of \citealt{gru12}]{gru09}. However, the situation appears somewhat different for $\theta^1$\,Ori\,C. While the absorption in \siiv\,$\lambda$\,1400 increases when the magnetic equator is seen edge-on, the \civ\ and \nv\ lines display an opposite behaviour compared to Of?p stars \citep{smi05}: these lines have enhanced blueshifted absorption when \ha\ emission is minimum (hence equator is seen edge-on) and enhanced absorption at lower velocities when \ha\ emission is maximum (face-on view of the equatorial regions). The reason for this opposite behaviour is not yet known.

\subsubsection{Modelling of the UV spectrum}

\begin{figure*}
  \begin{minipage}{8.0cm}
    \resizebox{\hsize}{!}  {\includegraphics[angle=90]{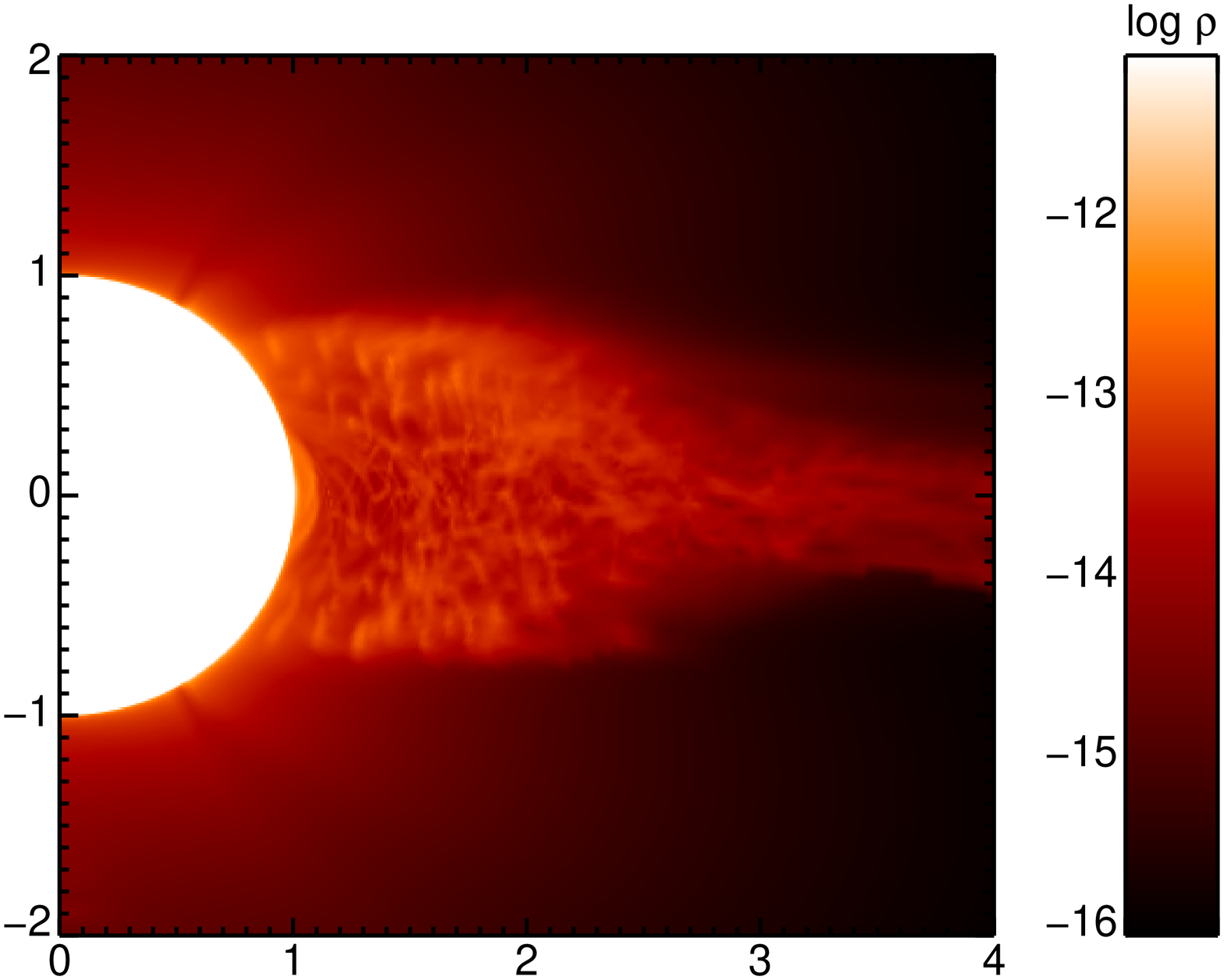}}
    \centering
   \end{minipage}
    \begin{minipage}{8.0cm}
    \resizebox{\hsize}{!}{\includegraphics[angle=90]{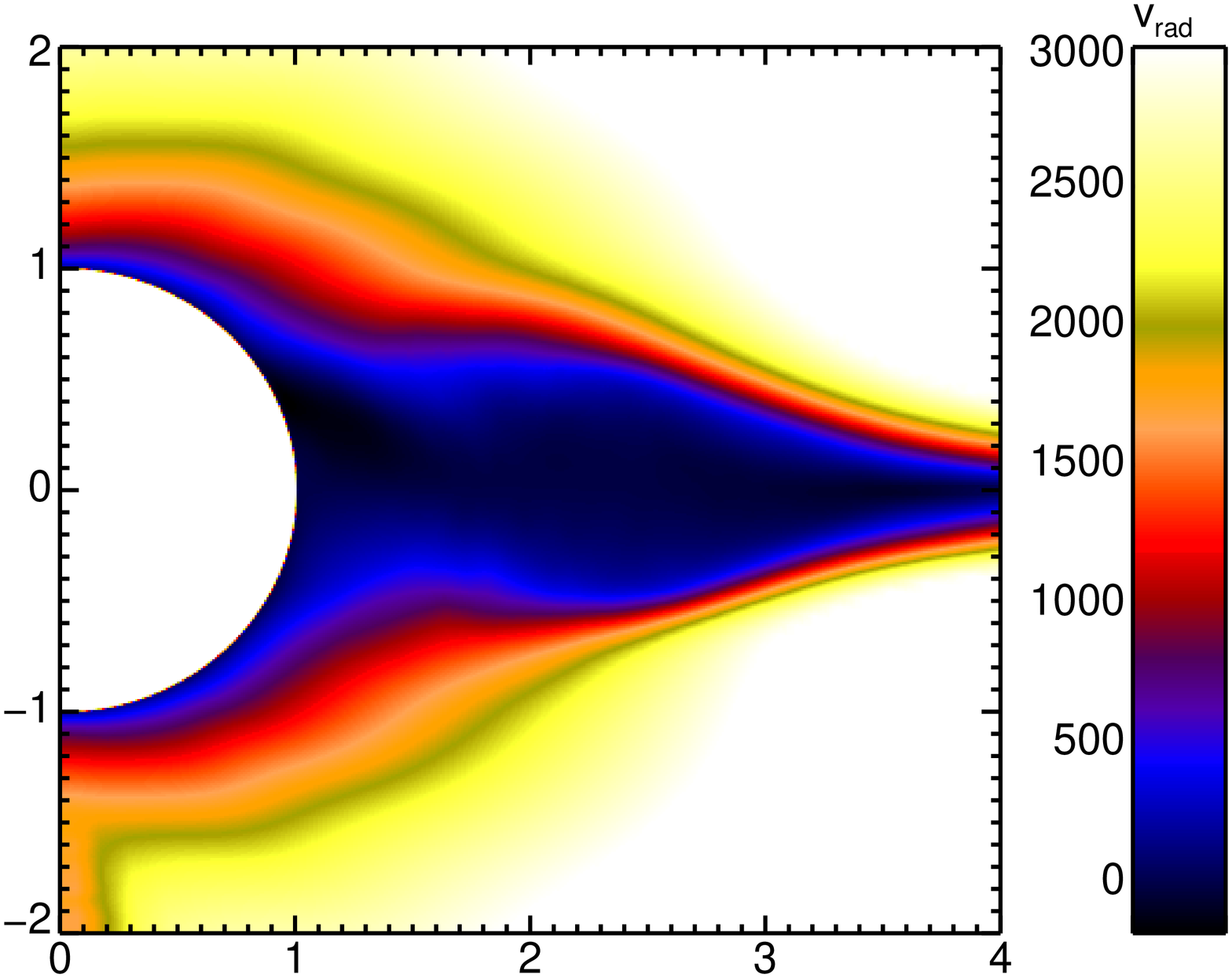}}
            \centering
   \end{minipage}         
  \caption{Plots of logarithmic density (left) in g\,cm$^{-3}$ and radial velocity (right) in km\,s$^{-1}$, computed by averaging 122 zones in azimuth for a snapshot of the 3-D radiation MHD wind simulation described in the text. The x- and y-axes are both scaled to units of $R_*$, and the magnetic north pole is located to the top at $x=0$.}
  \label{Fig:MHD}
\end{figure*}

We now interpret the UV observations described in the previous subsections by considering the structure of the magnetically confined winds in \cpd, for which $\eta_*$ amounts to $\sim$230 and $R_{\rm A}$ to $\sim 4.2\,R_*$ (Table \ref{prop}). We use a snapshot of a full 3-D numerical radiation magnetohydrodynamic (MHD) simulation that follows the dynamics of an O-star wind possessing a large-scale dipolar magnetic field, hence a large $\eta_*$ of $\approx 100$, as described in detail by \citet[see also Fig. \ref{Fig:MHD}]{udDoula13}. This model was initially calculated for HD\,191612 but since its properties are similar to those of \cpd, the model can directly be applied to the latter star (taking into account the uncertainties, see Table \ref{prop}). A key aspect of the model is the over-dense region around the magnetic equator, which is characterized by quite low velocities when compared to the free-flowing wind of a non-magnetic O star. Considering the geometry of the specific system under investigation, this general model of a dynamical magnetosphere has been able to reproduce the rotational modulation of the \ha\ line of \cpd\ \citep{wad15}, but also of other magnetic O-stars \citep{Sundqvist12, gru12, udDoula13, pet13}. Note however that, in addition to the magnetically confined plasma in the closed equatorial loops, the strong UV wind lines studied here depend also on the wind outflow in the open field regions \citep[see Fig. \ref{cartoon} and ][]{mar13}. 

To compute synthetic UV resonance line-profiles from our MHD simulation,  we describe the opacity by the parameter \citep[e.g.,][]{Puls93}:  
\begin{equation} 
  k_0 = \frac{q \, \dot{M}_{B=0}}{R_* v_\infty^2} \frac{\pi
    e^2/m_{\rm e}c}{4 \pi m_{\rm H}} \frac{A_{\rm i}}{1+4Y_{\rm
      He}} f_{\rm lu} \lambda_0.
\label{Eq:k0}
\end{equation} 
In this equation, $Y_{\rm He}$ is the helium number abundance, $f_{\rm lu}$ the oscillator strength of the transition, $\lambda_0$ its rest wavelength, $q$ the ion fraction of the considered element, $A_{\rm i} = n_{\rm i}/n_{\rm H}$ its abundance with respect to hydrogen, and the other symbols have their conventional meaning. 

An NLTE two-level atom scattering source function is then computed using the 3-D, local Sobolev method described, e.g., by \citet{cra96}. Finally, we solve the formal solution of radiative transfer in a 3-D cylindrical system for an observer viewing the magnetic axis with an angle $\alpha$ (see Table \ref{journal} and Fig. \ref{cartoon}), following \citet{Sundqvist12}. 

\begin{figure}
    \resizebox{\hsize}{!}  {\includegraphics[angle=90]{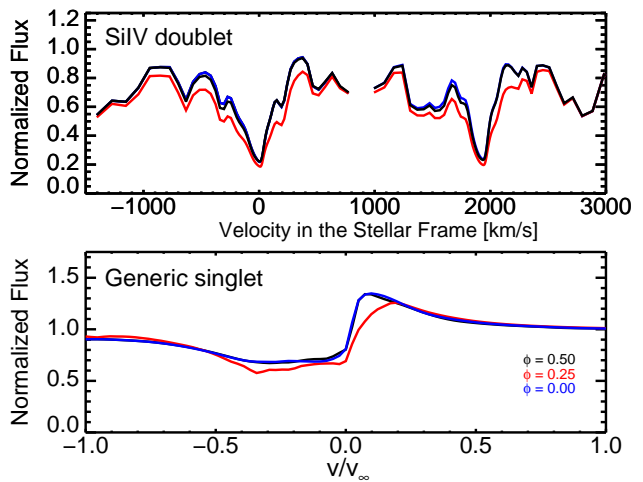}}
   \centering      
  \caption{Synthetic UV line-spectra for the three observed phases. The upper panel displays computations for the {\ion{Si}{4}} doublet, using $k_0 = 0.1$, while the lower panel shows a generic singlet profile with $k_0 = 1$, illustrative of the \civ\ and \nv\ lines. See text for discussion and model description. }
  \label{Fig:profs}
\end{figure}

The top panel of Fig.~\ref{Fig:profs} displays the synthetic UV line profiles for the (clearly separated) lines of the {\ion{Si}{4}} doublet. We have here used photospheric profiles from a {\sc tlusty} model atmosphere \citep{Lanz03}, for the stellar parameters in Table \ref{prop} and convolution by an isotropic 40 km\,s$^{-1}$ macroturbulence \citep{Sundqvist13}, as a lower boundary condition to our radiative transfer computations. However, our code cannot treat overlapping resonance doublets, such as {\ion{C}{4}} and {\ion{N}{5}}: neglecting the underlying photospheric profiles, these lines are instead discussed in terms of a generic singlet line of approximately the same observed line strength.\\

\vspace*{-0.1cm}

\noindent \textbf{Line strengths and ion balance.} The observed wind lines are surprisingly weak for an O star corresponding to the parameters given in Table \ref{prop}. The {\ion{Si}{4}} lines require an opacity parameter $k_0 \la 0.1$ in order that the emission scattering component does not become too strong. In addition, and perhaps more importantly, both the {\ion{C}{4}} and {\ion{N}{5}} P~Cygni lines are unsaturated. While the faster-than-radial divergence of the magnetic polar wind \citep{Owocki04} results in line profiles that are weaker than in corresponding non-magnetic winds (see \citealt{mar13}, their Fig. 6), test computations show that both the absorption trough and emission peak of the \civ\ and \nv\ lines become too strong when $k_0$ is increased by factors of a few above unity. Inserting  in eqn.~\ref{Eq:k0} appropriate line strengths (as used in Fig.~\ref{Fig:profs}) and other relevant parameters, and adopting a solar Si abundance and N and C abundances as estimated by \citet{wad15}, give for the product of ionization fraction and mass feeding rate, $\langle q \dot{M}_{B=0} \rangle \approx 0.4,\,6,\,9 \times 10^{-9} \, \rm M_\odot/yr$ for {\ion{Si}{4}}, {\ion{C}{4}}, {\ion{N}{5}}, respectively. Assuming then $\dot{M}_{B=0}$ according to the scaling formula by \citet[ see Table \ref{prop}]{Vink00}, this gives order-of-magnitude estimates for ion fractions $\langle q \rangle$ of about $\sim$0.01 for {\ion{C}{4}} and {\ion{N}{5}}, and $\la 0.001$ for {\ion{Si}{4}}. 

Unfortunately, as of today there exists no proper NLTE code for calculating the ion balance in a non-spherical, magnetic O-star wind, where most likely the distribution of X-rays also is highly aspherical (see previous sections). As a first approximation, however, we have computed ionization models by means of the spherically symmetric NLTE {\sc fastwind} code \citep{Puls05}, using a newly implemented treatment of X-rays (Carneiro et al., in prep., essentially following \citealt{Pauldrach94} and \citealt{Feldmeier97}). Fig. ~\ref{Fig:ion} displays the computed ion fractions at a characteristic radius $r = 2 R_*$, for three different models with $L_{\rm x}/L_{\rm Bol} $ ratios in the 0.4-2.5\,keV energy band of 0, $\sim 10^{-7}$, and $\sim 10^{-6}$. 

Focusing on the well developed -- but unsaturated --  {\ion{C}{4}} and {\ion{N}{5}} wind lines, the figure shows that without any X-ray ionization the {\ion{C}{4}} (resp. {\ion{N}{5}}) ion fractions are far too high (resp. low) compared to observationally-derived values. Introducing $L_{\rm x}/L_{\rm Bol} \sim 10^{-7}$ does not significantly improve the situation, since {\ion{C}{4}} remains the main ion stage. As the X-ray luminosity increases,  {\ion{C}{4}} starts to be ionized away, whereas the fraction of {\ion{N}{5}} steadily increases. Since the {\ion{C}{4}} and {\ion{N}{5}} lines are unsaturated and of similar strength, this anti-correlated behaviour of their ion fractions is a potentially important clue for understanding the strength of the UV wind lines in \cpd\ and similar magnetic stars. 

Since the \civ\ and \nv\ lines are observed to be of approximately equal strengths, their ion fractions must also be roughly equal. From Fig. \ref{Fig:ion}, we see this occurs in our models for a $L_{\rm x}/L_{\rm Bol}\sim 10^{-6.3}$, significantly higher than the canonical value of $10^{-7}$ for non-magnetic O-stars \citep{ber97,naz09,naz11}, in general agreement but still slightly below the observed X-ray luminosity for \cpd\ (\loglxlb\ $\sim -6.1$ in that bandpass, see models of Section 3). The \siiv\ ion fraction is then $q \sim0.001$, as derived above, but the {\ion{C}{4}} and {\ion{N}{5}} ion fractions are $q \sim 0.1$, while we instead derived values of $\sim$0.01, assuming a mass-loss feeding rate from Vink's formula. Matching the {\ion{C}{4}} and {\ion{N}{5}} line profiles with $q \sim 0.1$ would require the $\dot{M}_{B=0}$ assumed above be reduced by approximately an order of magnitude. This result is similar to that found by \citet{mar13} for HD\,191612, but it should be kept in mind that the ion fractions presented here are valid only at a certain characteristic radius whereas the real $q$ in the non-spherical, magnetic wind will certainly vary both with radius and latitude. As such, these tentative results need to be tested by more direct comparisons to the observed line profiles, using a new, future set of improved models.  \\

\begin{figure}
    \resizebox{\hsize}{!}  {\includegraphics[angle=90]{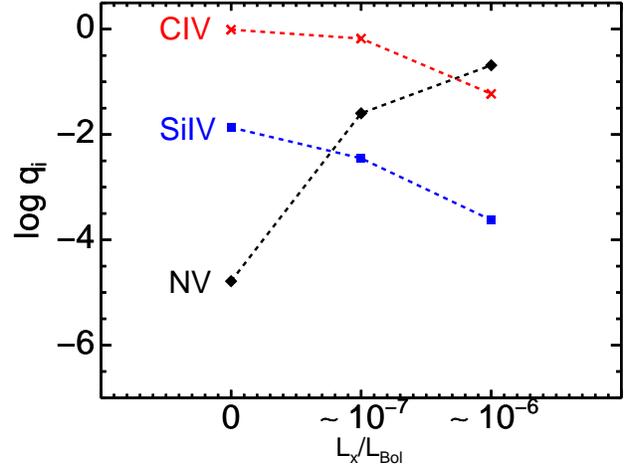}}
   \centering      
  \caption{Theoretical calculations of the logarithms of ion fractions $q_{\rm i}$ for $i = $ {\ion{C}{4}}, {\ion{Si}{4}}, and {\ion{N}{5}} (labelled in the figure) at a characteristic radius $r =2R_*$ in {\sc fastwind} models using three different $L_{\rm x}/L_{\rm Bol}$ ratios (see text for details).}
  \label{Fig:ion}
\end{figure}

\vspace*{-0.1cm}

\noindent \textbf{Variability.} As discussed by \citet{mar13}, the line-of-sight absorption column when viewed from above the magnetic equator samples the high density wind material of the magnetosphere, which is characterized by rather low velocities and some infall. By contrast, the absorption column from an observer above the magnetic pole samples the free-streaming and fast polar wind. This leads to stronger absorption at high (resp. low) velocities when viewed closer to the magnetic pole (resp. equator). Even though we never view \cpd\ from directly above the magnetic pole (Table \ref{journal} and Fig. \ref{cartoon}), this quite general behavior is preserved. It explains the stronger absorptions seen at low velocities for $\phi$=0.25 in the {\ion{He}{2}}, {\ion{Si}{4}}, and {\ion{C}{4}} lines (Figs. \ref{specuv} and \ref{uv_spectra2}), while the absorption difference at high velocities between the $\phi$=0.25 (equator-on) and $\phi$=0.5 views is reduced in the generic line profile (bottom panel of Fig. \ref{Fig:profs} to be compared with third panel of Fig. \ref{uv_spectra2}). However, the observed weak silicon lines are also highly influenced by the interstellar absorption and photospheric profile. In our models, the latter lowers the scattering emission component and brings these lines into pure absorption. Overall, our synthetic line profiles and their variability are in good qualitative agreement with the observed profiles in Fig.~\ref{uv_spectra2}, providing further strong support for the general ``dynamical magnetosphere'' paradigm for magnetized O-star winds. 

There are some potentially important deviations between models and observations, however. For example, the change in width of the \siiv\ line is quantitatively larger in the STIS observations. In addition, the large difference in absorption at high velocities in the {\ion{C}{4}} and {\ion{N}{5}} lines during phases 0.0 and 0.5 indicates strong asymmetries about the magnetic equator. While transient asymmetries are seen in our 3D MHD simulations \citep[see also][]{udDoula13}, their corresponding effect on the synthetic line profiles in Fig.~\ref{Fig:profs} is barely visible and as such is much weaker than observed. It is interesting to note that these observed absorption differences between equal viewing angles toward the southern and northern hemispheres occur mainly at high blueshifted velocities, suggesting (somewhat surprisingly) that the free streaming wind might be more affected by such symmetry-breaking than the confined dynamical magnetosphere.
 
\section{Conclusion}

We have examined the high-energy emission of the Of?p star \cpd, and its variations, using \xmm\ and \hst\ observations. In this system, the magnetic field is able to confine the stellar wind flows near the magnetic equator, forming a dynamical magnetosphere surrounding the star. With angles of inclination and obliquity of (90$^{\circ}$, 35$^{\circ}$) or (35$^{\circ}$, 90$^{\circ}$), the magnetosphere of \cpd\ is seen edge-on twice per rotation period (corresponding to minimum emissions in the optical) and nearly face-on twice per period (corresponding to maximum emissions in the optical). The \xmm\ observations sample both maxima and minima, as well as an intermediate phase; the \hst\ observations sample both maxima and the first minimum.

In X-rays, \cpd\ displays a bright (\loglxlb $\sim -5.8$) and hard (hard-to-soft flux ratio close to one) emission. The X-ray emission is both brighter and harder than in other Of?p stars. This emission remains stable on short timescales, but the X-ray flux varies by $\sim$55\% with rotational phase, a value similar to what is seen in the Of?p star HD\,191612 or in the O-star $\theta^1$\,Ori\,C. These phase-locked changes closely follow the variations of the optical emission lines, i.e. there are two maxima and two minima in X-rays during the 73d rotational period of \cpd, as expected for a magnetic oblique rotator. There is no significant hardness variation except for the last observation, taken during the second minimum. In view of the stability of the behaviour in the optical domain, this change is probably linked to an asymmetry in the magnetosphere.

In the UV domain, two types of data are available: photometry taken by the OM telescope aboard \xmm\ and high-quality \hst-STIS spectroscopy. These observations reveal that \cpd\ displays a stable broad-band flux as well as stable ``photospheric'' lines. However, large profile variations of the lines associated with the circumstellar environment are detected in the UV spectra. First, enhanced absorption at low velocities is observed when the magnetic equatorial regions are seen edge-on. This increase in absorption, particularly spectacular for the {\ion{He}{2}\,$\lambda$\,1640} line, is directly due to the presence of dense plasma projected onto the stellar disk. It is reproduced qualitatively by detailed 3D modelling of a magnetically confined wind. Second, a difference also exists in the high-velocity absorption of the {\ion{C}{4}} and {\ion{N}{5}} P~Cygni profiles when comparing the two phases corresponding to the two maxima of the optical emissions. This strong variation in profile appears surprising as the optical emissions at the same phases have similar strengths, but it nevertheless suggests the presence of asymmetries in the north vs the south magnetic hemispheres. As the difference appears at high velocities while the confined winds have low velocities, these asymmetries should be more prevalent in the free-flowing wind. However, the origin of these asymmetries and their link with the magnetic field remains to be established. Finally, we note that empirically derived ion fractions require a significantly higher $L_{\rm x}/L_{\rm Bol}$ ratio than the canonical value $10^{-7}$ for non-magnetic O-stars, in agreement with the detailed X-ray analysis, but that reproducing the overall strengths of the UV wind lines with these ion fractions then requires a $\dot{M}_{B=0}$ that is significantly lower than expected.

\section*{Acknowledgments}
YN \& GR acknowledge support from  the Fonds National de la Recherche Scientifique (Belgium), the PRODEX XMM contract, and an ARC grant for Concerted Research Action financed by the Federation Wallonia-Brussels. JOS acknowledges support from SAO Chandra grant TM3-14001A. AuD acknowledges support by NASA through Chandra Award number TM4-15001A and 16200111 issued by the Chandra X-ray Observatory Center which is operated by the Smithsonian Astrophysical Observatory for and behalf of NASA under contract NAS8- 03060. AuD, AWF, and NRW acknowledge support for Program number HST-GO-13629 that was provided by NASA through a grant from the Space Telescope Science Institute. GAW acknowledges Discovery Grant support from the Natural Science and Engineering Research Council (NSERC) of Canada. The authors also acknowledge help and discussion with Rodolfo Barb\`a and V\'eronique Petit. We thank Luiz Carneiro for making a new version of FASTWIND available to us prior to final publication. ADS and CDS were used for preparing this document. 


\bsp

\label{lastpage}


\begin{thebibliography}{}
\bibitem[Asplund et al.(2009)]{aspl} Asplund, M., Grevesse, N., Sauval, A.~J., \& Scott, P.\ 2009, ARA\&A, 47, 481 
\bibitem[Babel \& Montmerle(1997a)]{bab97} Babel, J., \& Montmerle, T.\ 1997a, A\&A, 323, 121
\bibitem[Barannikov(2007)]{bar07} Barannikov, A.~A.\ 2007, Information Bulletin on Variable Stars, 5756, 1 
\bibitem[Barb{\'a} et al.(2010)]{bar10} Barb{\'a}, R.~H., Gamen, R., Arias, J.~I., et al.\ 2010, Revista Mexicana de Astronomia y Astrofisica Conference Series, 38, 30 
\bibitem[Barb{\'a} et al.(2014)]{bar14} Barb{\'a}, R., Gamen, R., Arias, J.~I., et al.\ 2014, Revista Mexicana de Astronomia y Astrofisica Conference Series, 44, 148 
\bibitem[Berghoefer et al.(1997)]{ber97} Berghoefer, T.~W., Schmitt, J.~H.~M.~M., Danner, R., \& Cassinelli, J.~P.\ 1997, A\&A, 322, 167 
\bibitem[Bohlin et al.(1978)]{boh78} Bohlin, R.~C., Savage, B.~D., \& Drake, J.~F.\ 1978, ApJ, 224, 132
\bibitem[Cranmer \& Owocki(1996)]{cra96} Cranmer, S.~R., \& Owocki, S.~P.\ 1996, ApJ, 462, 469 
\bibitem[Donati et al.(2006)]{don06} Donati, J.-F., Howarth, I.~D., Jardine, M.~M., et al.\ 2006, MNRAS, 370, 629 
\bibitem[{{Feldmeier} {et~al}\mbox{.}(1997){Feldmeier}, {Kudritzki}, {Palsa},  {Pauldrach}, \& {Puls}}]{Feldmeier97} {Feldmeier} A., {Kudritzki} R.-P., {Palsa} R., {Pauldrach} A.~W.~A., {Puls} J., 1997, A\&A, 320, 899
\bibitem[Gagn{\'e} et al.(2005)]{gag05} Gagn{\'e}, M., Oksala, M.~E., Cohen, D.~H., et al.\ 2005, ApJ, 628, 986 (see erratum in ApJ, 634, 712)
\bibitem[Garrison et al.(1977)]{gar77} Garrison, R.~F., Hiltner, W.~A., \& Schild, R.~E.\ 1977, ApJS, 35, 111 
\bibitem[Grunhut et al.(2009)]{gru09} Grunhut, J.~H., Wade, G.~A., Marcolino, W.~L.~F., et al.\ 2009, MNRAS, 400, L94 
\bibitem[Grunhut et al.(2012)]{gru12} Grunhut, J.~H., Wade, G.~A., Sundqvist, J.~O., et al.\ 2012, MNRAS, 426, 2208 
\bibitem[Hubrig et al.(2011)]{hub11} Hubrig, S., Sch{\"o}ller, M., Kharchenko, N.~V., et al.\ 2011, A\&A, 528, AA151 
\bibitem[Hubrig et al.(2012)]{hub12} Hubrig, S., Kholtygin, A., Scholler, M., et al.\ 2012, Information Bulletin on Variable Stars, 6019, 1 
\bibitem[Hubrig et al.(2015)]{hub14} Hubrig, S., Sch{\"o}ller, M., Kholtygin, A.~F., et al.\ 2015, MNRAS, 447, 1885 
\bibitem[Jansen et al.(2001)]{jan01} Jansen, F., Lumb, D., Altieri, B., et al.\ 2001, A\&A, 365, L1 
\bibitem[{{Lanz} \& {Hubeny}(2003)}]{Lanz03} {Lanz} T., {Hubeny} I., 2003, ApJS, 146, 417
\bibitem[Marcolino et al.(2012)]{mar12} Marcolino, W.~L.~F., Bouret, J.-C., Walborn, N.~R., et al.\ 2012, MNRAS, 422, 2314 
\bibitem[Marcolino et al.(2013)]{mar13} Marcolino, W.~L.~F., Bouret, J.-C., Sundqvist, J.~O., et al.\ 2013, MNRAS, 431, 2253 
\bibitem[Martins et al.(2010)]{mar10} Martins, F., Donati, J.-F., Marcolino, W.~L.~F., et al.\ 2010, MNRAS, 407, 1423 
\bibitem[Mason et al.(2001)]{mas01} Mason, K.~O., Breeveld, A., Much, R., et al.\ 2001, A\&A, 365, L36 
\bibitem[Naz{\'e}(2009)]{naz09} Naz{\'e}, Y.\ 2009, A\&A, 506, 1055 
\bibitem[Naz\'e et al.(2001)]{naz01} Naz\'e, Y., Vreux, J.-M., Rauw, G. 2001, A\&A 372, 195
\bibitem[Naz\'e et al.(2004)]{naz04} Naz\'e, Y., Rauw, G., Vreux, J.-M., \& De Becker, M.\ 2004, A\&A, 417, 667
\bibitem[Naz\'e et al.\ (2006)]{naz06} Naz\'e, Y., Barbieri, C., Segafredo, A., Rauw, G., \& De Becker, M.\ 2006, IBVS, 5693
\bibitem[Naz{\'e} et al.(2007)]{naz07} Naz{\'e}, Y., Rauw, G., Pollock, A.~M.~T., Walborn, N.~R., \& Howarth, I.~D.\ 2007, MNRAS, 375, 145 
\bibitem[Naz\'e et al.(2008a)]{naz08} Naz\'e, Y., Walborn, N.R., Rauw, G., Martins, F., Pollock, A.M.T., \& Bond, H.E.\ 2008a, AJ, 135, 1946
\bibitem[Naz{\'e} et al.(2008b)]{naz08b} Naz{\'e}, Y., Walborn, 
N.~R., \& Martins, F.\ 2008b, Revista Mexicana de Astronomia y Astrofisica, 44, 331 
\bibitem[Naz{\'e} et al.(2010)]{naz10} Naz{\'e}, Y., ud-Doula, A., Spano, M., et al.\ 2010, A\&A, 520, A59 
\bibitem[Naz{\'e} et al.(2011)]{naz11} Naz{\'e}, Y., Broos, P.~S., Oskinova, L., et al.\ 2011, ApJS, 194, 7 
\bibitem[Naz{\'e} et al.(2013)]{naz13} Naz{\'e}, Y., Oskinova, L.~M., \& Gosset, E.\ 2013, ApJ, 763, 143 
\bibitem[Naz{\'e} et al.(2014a)]{naz14} Naz{\'e}, Y., Petit, V., Rinbrand, M., et al.\ 2014a, ApJS, 215, 10 
\bibitem[Naz{\'e} et al.(2014b)]{naz14b} Naz{\'e}, Y., Wade, G.~A., \& Petit, V.\ 2014b, A\&A, 569, A70 
\bibitem[{{Owocki} \& {ud-Doula}(2004)}]{Owocki04}{Owocki} S.~P., {ud-Doula} A., 2004, ApJ, 600, 1004
\bibitem[{{Pauldrach} {et~al}\mbox{.}(1994){Pauldrach}, {Kudritzki}, {Puls},  {Butler}, \& {Hunsinger}}]{Pauldrach94} {Pauldrach} A.~W.~A., {Kudritzki} R.~P., {Puls} J., {Butler} K., {Hunsinger}  J., 1994, A\&A, 283, 525
\bibitem[Petit et al.(2013)]{pet13} Petit, V., Owocki, S.~P., Wade, G.~A., et al.\ 2013, MNRAS, 429, 398 
\bibitem[{{Puls}, {Owocki} \& {Fullerton}(1993){Puls}, {Owocki}, \&  {Fullerton}}]{Puls93}{Puls} J., {Owocki} S.~P., {Fullerton} A.~W., 1993, A\&A, 279, 457
\bibitem[{{Puls} {et~al}\mbox{.}(2005){Puls}, {Urbaneja}, {Venero}, {Repolust},
  {Springmann}, {Jokuthy}, \& {Mokiem}}]{Puls05}{Puls} J., {Urbaneja} M.~A., {Venero} R., {Repolust} T., {Springmann} U.,  {Jokuthy} A., {Mokiem} M.~R., 2005, A\&A, 435, 669
\bibitem[Smith \& Fullerton(2005)]{smi05} Smith, M.~A., \& Fullerton, A.~W.\ 2005, PASP, 117, 13 
\bibitem[Stahl et al.(1996)]{sta96} Stahl, O., Kaufer, A., Rivinius, T., et al.\ 1996, A\&A, 312, 539 
\bibitem[Stelzer et al.(2005)]{ste05} Stelzer, B., Flaccomio, E., Montmerle, T., et al.\ 2005, ApJS, 160, 557
\bibitem[Str{\"u}der et al.(2001)]{str01} Str{\"u}der, L., Briel, U., Dennerl, K., et al.\ 2001, A\&A, 365, L18 
\bibitem[{{Sundqvist} {et~al}\mbox{.}(2013){Sundqvist}, {Petit}, {Owocki},  {Wade}, {Puls}, \& {MiMeS Collaboration}}]{Sundqvist13}{Sundqvist} J.~O., {Petit} V., {Owocki} S.~P., {Wade} G.~A., {Puls} J., {MiMeS  Collaboration}, 2013, MNRAS, 433, 2497
\bibitem[{{Sundqvist} {et~al}\mbox{.}(2012){Sundqvist}, {ud-Doula}, {Owocki},  {Townsend}, {Howarth}, \& {Wade}}]{Sundqvist12} {Sundqvist} J.~O., {ud-Doula} A., {Owocki} S.~P., {Townsend} R.~H.~D.,  {Howarth} I.~D., {Wade} G.~A., 2012, MNRAS, 423, L21
\bibitem[{{ud-Doula} \& {Owocki}(2002)}]{udDoula02} {ud-Doula} A., {Owocki} S.~P., 2002, ApJ, 576, 413
\bibitem[{{ud-Doula}, {Owocki} \& {Townsend}(2008){ud-Doula}, {Owocki}, \&  {Townsend}}]{udDoula08} {ud-Doula} A., {Owocki} S.~P., {Townsend} R.~H.~D., 2008, MNRAS, 385, 97
\bibitem[ud-Doula et al.(2009)]{udd09} ud-Doula, A., Owocki, S.~P., \& Townsend, R.~H.~D.\ 2009, MNRAS, 392, 1022 
\bibitem[{{ud-Doula} {et~al}\mbox{.}(2013){ud-Doula}, {Sundqvist}, {Owocki},  {Petit}, \& {Townsend}}]{udDoula13} {ud-Doula} A., {Sundqvist} J.~O., {Owocki} S.~P., {Petit} V., {Townsend}  R.~H.~D., 2013, MNRAS, 428, 2723
\bibitem[ud-Doula et al.(2014)]{udd14} ud-Doula, A., Owocki, S., Townsend, R., Petit, V., \& Cohen, D.\ 2014, MNRAS, 441, 3600 
\bibitem[Turner et al.(2001)]{tur01} Turner, M.~J.~L., Abbey, A., Arnaud, M., et al.\ 2001, A\&A, 365, L27 
\bibitem[{{Vink}, {de Koter} \& {Lamers}(2000){Vink}, {de Koter}, \&  {Lamers}}]{Vink00} {Vink} J.~S., {de Koter} A., {Lamers} H.~J.~G.~L.~M., 2000, A\&A, 362, 295
\bibitem[Wade et al.(2015)]{wad15} Wade, G.~A., Barb{\'a}, R.~H., Grunhut, J., et al.\ 2015, MNRAS, 447, 2551
\bibitem[Walborn(1972)]{wal72} Walborn, N.R. 1972, AJ, 77, 312
\bibitem[Walborn(1973)]{wal73} Walborn N.R. 1973, AJ, 78, 1067
\bibitem[Walborn et al.(2000)]{wal00} Walborn, N.~R., Lennon, D.~J., Heap, S.~R., et al.\ 2000, PASP, 112, 1243 
\bibitem[Walborn et al.(2004)]{wal04} Walborn, N.R., Howarth, I.D., Rauw, G., et al. 2004, ApJ, 617, L61 
\bibitem[Walborn et al.(2010)]{wal10} Walborn, N.~R., Sota, A., Ma{\'{\i}}z Apell{\'a}niz, J., et al.\ 2010, ApJ, 711, L143 
\end{thebibliography}
\end{document}